\PassOptionsToPackage{unicode}{hyperref}
\PassOptionsToPackage{hyphens}{url}
\documentclass[
]{article}
\usepackage{xcolor}
\usepackage[margin=1in]{geometry}
\usepackage{amsmath,amssymb}
\usepackage{iftex}
\ifPDFTeX
  \usepackage[T1]{fontenc}
  \usepackage[utf8]{inputenc}
  \usepackage{textcomp} % provide euro and other symbols
\else % if luatex or xetex
  \usepackage{unicode-math} % this also loads fontspec
  \defaultfontfeatures{Scale=MatchLowercase}
  \defaultfontfeatures[\rmfamily]{Ligatures=TeX,Scale=1}
\fi
\usepackage{lmodern}
\ifPDFTeX\else
\fi
\IfFileExists{upquote.sty}{\usepackage{upquote}}{}
\IfFileExists{microtype.sty}{% use microtype if available
  \usepackage[]{microtype}
  \UseMicrotypeSet[protrusion]{basicmath} % disable protrusion for tt fonts
}{}
\usepackage{setspace}
\makeatletter
\@ifundefined{KOMAClassName}{% if non-KOMA class
  \IfFileExists{parskip.sty}{%
    \usepackage{parskip}
  }{% else
    \setlength{\parindent}{0pt}
    \setlength{\parskip}{6pt plus 2pt minus 1pt}}
}{% if KOMA class
  \KOMAoptions{parskip=half}}
\makeatother
\usepackage{color}
\usepackage{fancyvrb}

\DefineVerbatimEnvironment{Highlighting}{Verbatim}{commandchars=\\\{\}}
\usepackage{framed}
\definecolor{shadecolor}{RGB}{248,248,248}
\newenvironment{Shaded}{\begin{snugshade}}{\end{snugshade}}

\newcommand{\CommentTok}[1]{\textcolor[rgb]{0.56,0.35,0.01}{\textit{#1}}}

\newcommand{\ControlFlowTok}[1]{\textcolor[rgb]{0.13,0.29,0.53}{\textbf{#1}}}
\newcommand{\DataTypeTok}[1]{\textcolor[rgb]{0.13,0.29,0.53}{#1}}
\newcommand{\DecValTok}[1]{\textcolor[rgb]{0.00,0.00,0.81}{#1}}

\newcommand{\FloatTok}[1]{\textcolor[rgb]{0.00,0.00,0.81}{#1}}

\newcommand{\KeywordTok}[1]{\textcolor[rgb]{0.13,0.29,0.53}{\textbf{#1}}}
\newcommand{\NormalTok}[1]{#1}

\usepackage{graphicx}
\makeatletter
\newsavebox\pandoc@box
\newcommand*\pandocbounded[1]{% scales image to fit in text height/width
  \sbox\pandoc@box{#1}%
  \Gscale@div\@tempa{\textheight}{\dimexpr\ht\pandoc@box+\dp\pandoc@box\relax}%
  \Gscale@div\@tempb{\linewidth}{\wd\pandoc@box}%
  \ifdim\@tempb\p@<\@tempa\p@\let\@tempa\@tempb\fi% select the smaller of both
  \ifdim\@tempa\p@<\p@\scalebox{\@tempa}{\usebox\pandoc@box}%
  \else\usebox{\pandoc@box}%
  \fi%
}
\def\fps@figure{htbp}
\makeatother
\NewDocumentCommand\citeproctext{}{}

\makeatletter
 \let\@cite@ofmt\@firstofone
 \def\@biblabel#1{}
 \def\@cite#1#2{{#1\if@tempswa , #2\fi}}
\makeatother
\newlength{\cslhangindent}
\newlength{\csllabelwidth}
\newenvironment{CSLReferences}[2] % #1 hanging-indent, #2 entry-spacing
 {\begin{list}{}{%
  \setlength{\itemindent}{0pt}
  \setlength{\leftmargin}{0pt}
  \setlength{\parsep}{0pt}
  \ifodd #1
   \setlength{\leftmargin}{\cslhangindent}
   \setlength{\itemindent}{-1\cslhangindent}
  \fi
  \setlength{\itemsep}{#2\baselineskip}}}
 {\end{list}}
\usepackage{calc}

\usepackage[hang,flushmargin]{footmisc}
\usepackage{titling}
\usepackage{booktabs}
\usepackage{longtable}
\usepackage{array}
\usepackage{multirow}
\usepackage{wrapfig}
\usepackage{float}
\usepackage{colortbl}
\usepackage{pdflscape}
\usepackage{tabu}
\usepackage{threeparttable}
\usepackage{threeparttablex}
\usepackage[normalem]{ulem}
\usepackage{makecell}
\usepackage{xcolor}
\usepackage{bookmark}
\IfFileExists{xurl.sty}{\usepackage{xurl}}{} % add URL line breaks if available
\hypersetup{
  hidelinks,
  pdfcreator={LaTeX via pandoc}}

\title{Estimating the Number of Street Vendors in New York City:\\
Ratio Estimation with Point Process Data\thanks{Thanks to Mohamed Attia, Eric Auerbach, Debipriya
Chatterjee, David Kallick, Carina Kaufman-Gutierrez, Joseph Salvo,\\
\strut ~~Shamier Settle, Matthew Shapiro, Emerita Torres, and Anand
Vidyashankar for discussion and feedback on the work described\\
\strut ~~in this paper.} \vspace{1em}}
\author{Jonathan Auerbach\\
Department of Statistics\\
George Mason University\\
\href{mailto:jauerba@gmu.edu}{\nolinkurl{jauerba@gmu.edu}}}
\date{\vspace{-1em}}

\begin{document}
\maketitle

\setstretch{1.5}
\small

\begin{center}
\textbf{Abstract}
\end{center}

\vspace{1em}

\begin{center}
\begin{minipage}{.9\linewidth}
We estimate the number of street vendors in New York City. First, we summarize the process by which vendors receive licenses and permits to operate legally in New York City. We then describe a survey that was administered by the Street Vendor Project while distributing coronavirus relief aid to vendors operating in New York City both with and without a license or permit. Finally, we review ratio estimation and develop a theoretical justification based on the theory of point processes. We find approximately 23,000 street vendors operate in New York City---20,500 mobile food vendors and 2,400 general merchandise vendors---with one third located in just six ZIP Codes---11368 (16\%), 11372 (3\%), and 11354 (3\%) in North and West Queens and 10036 (5\%), 10019 (4\%), and 10001 (3\%) in the Chelsea and Clinton neighborhoods of Manhattan. Our estimates suggest the American Community Survey misses the majority of New York City street vendors.
\end{minipage}
\end{center}

\vspace{1em}

\normalsize

\subsection{1. Introduction}\label{introduction}

Street vendors are New York City's smallest businesses, selling food and
merchandise from carts, stalls, and trucks throughout the five boroughs.
They are an iconic part of the urban landscape. They are also a thriving
sector of the local economy, contributing millions of dollars in
government revenue through taxes, fines, and fees. Perhaps most
importantly, street vending historically benefits underserved
communities, both because vendors operate in neighborhoods with limited
access to traditional stores and because vending is one of a handful of
occupations in which New Yorkers of all backgrounds, immigrants in
particular, can achieve economic mobility and a chance at the American
dream (Burrows and Wallace 1998, chap. 42).

Yet despite their social and economic importance, little is known about
the size and location of New York City's street vending population. This
is because while local law requires street vendors to obtain licenses
and permits to operate legally, the number of licenses and permits are
limited, resulting in a largely unknown population of vendors that
operate without a license or permit. These vendors are not easily
identified from administrative datasets, such as tax records or fines,
and they can be difficult to locate for government surveys, such as the
American Community Survey. Nevertheless, understanding the size and
location of New York's street vending population is crucial for
informing policy and advocacy.

In this paper, we propose and justify an estimator for the number of
street vendors in New York City, including those that operate without a
license or permit. We present our work in six sections:

In Section 2, we review the process by which vendors receive licenses
and permits, and we describe a survey administered by the Street Vendor
Project at the Urban Justice Center while distributing coronavirus
relief aid to vendors operating both with and without a license or
permit.

In Section 3, we use the fact that the number of licenses and permits
are limited by law to construct a ratio estimator for the number of
street vendors. We then develop a theoretical justification based on the
theory of point processes, which assumes the spatial distribution of
survey respondents is representative and well approximated by a set of
inhomogeneous Poisson processes. Finally, we derive the standard errors.

In Section 4, we examine the sensitivity of the ratio estimator to
deviations from the underlying assumptions. We first consider the
scenario in which the spatial distribution of the survey respondents is
not representative, and we examine a more general estimator in which
sampling weights adjust for differential nonresponse. We then consider
the scenario in which the Poisson process assumption is violated, and
the negative binomial process provides a better approximation. For both
scenarios, we rederive the standard errors.

In Section 5, we estimate the number of street vendors in New York City.
We find approximately 23,000 street vendors operate in New York City:
20,500 mobile food vendors and 2,400 general merchandise vendors. We
also find that this estimate is somewhat sensitive to deviations from
our underlying assumptions. For example, we find that if the spatial
distribution of survey respondents is not representative, the estimated
number of street vendors in New York City could be as low as 17,200 and
as high as 41,800.

In Section 6, we conclude with a discussion. We first consider how the
theoretical arguments provided in Sections 3 and 4 could be extended to
more complex models. We then consider some implications of the estimates
provided in Section 5. We find that although the spatial distribution of
respondents is consistent with the results of the American Community
Survey, the two disagree on the size of New York City's street vending
population. Our estimates suggest the American Community Survey misses
the majority of New York City street vendors. We believe these findings
shed light on the broader challenges of measuring the ``gig economy.''

\subsection{2. Background and Survey
Operations}\label{background-and-survey-operations}

A street vendor is any individual who sells goods from a mobile vending
unit instead of a store. We distinguish between the vending
establishment---the mobile vending unit from which the goods are
sold---and the vendors---the individuals who own and/or operate that
unit. Note that an establishment refers to a single vending unit. There
may be multiple vendors associated with any establishment, and multiple
establishments may be owned by a single individual or firm. For example,
suppose Sally owns the business Sally's Salads, which consists of two
food trucks, each staffed by three employees. Then Sally's Salads counts
as one business, two establishments, and seven vendors including Sally.

We limit our analysis to two types of street vendors: mobile food
vendors---vendors who sell food items such as sandwiches, drinks, and
cut fruit---and general merchandise vendors---vendors who sell
merchandise items such as electronics, clothing, and accessories. We
refer to these vendors as food vendors and merchandise vendors,
respectively. In Section 2.1, we review the process by which vendors
receive licenses and permits as required by law. In Section 2.2, we
describe a survey administered by the Street Vendor Project while
distributing coronavirus relief aid to vendors operating in New York
City both with and without a license or permit.

\subsubsection{2.1 Background}\label{background}

A long list of rules determines the individuals, locations, and time
periods during which vendors can legally sell their goods in New York
City. Most relevant is the fact that a merchandise vendor requires a
license to operate legally, and a food vendor requires both a license
and a permit. The number of food permits and merchandise licenses are
limited by law: 5,100 permits and 853 general merchandise licenses are
available at the time of our writing. Food licenses are not limited.
Note that a food permit is issued to an individual or business to allow
for the sale of food from a specific mobile food vending unit, such as a
cart or truck. A food license authorizes an individual to prepare or
serve food from a permitted mobile food vending unit.

There are several versions of a food permit with varying restrictions.
Of the 5,100 permits available, 200 are borough permits that limit
vendors to one of the five boroughs; 100 are reserved for veterans or
vendors with a disability; 1,000 are seasonal and valid only from April
to October; 1,000 are green cart permits that limit vendors to selling
fruit, vegetables, plain nuts, and water; and 2,800 are unrestricted.
Multiple individuals with food licenses can legally operate from one
permitted food vending unit. Merchandise licenses are renewed annually,
while food licenses and permits are renewed biennially.

There are three relevant exceptions to these rules. The first exception
is that merchandise vendors who are veterans are not subject to the
limit of 853 licenses. According to data obtained by Mosher and
Turnquist (2024), there are approximately 1,000 licensed merchandise
vendors who are veterans in New York City. We assume the number of
veterans selling merchandise without a license is negligible.

The second exception is First Amendment vendors. A First Amendment
vendor sells expressive merchandise such as newspapers, books, and art.
Expressive merchandise is considered free speech and protected under the
First Amendment of the U.S. Constitution so that the number of First
Amendment vendors cannot be restricted by law.

The final exception are concessionaires that operate on New York City
parkland through an NYC Parks license or permit. According to the New
York City Department of Parks and Recreation, there are approximately
400 concessions within New York City parks. Many offer food services,
ranging from food carts to restaurants. We do not consider First
Amendment vendors or park concessionaires in this paper. Our estimates
are limited to food and merchandise vendors operating outside of parks.

\subsubsection{2.2 Survey Operations}\label{survey-operations}

The Street Vendor Project (SVP), part of the Urban Justice Center, is a
non-profit organization that advocates on behalf of New York City street
vendors. SVP administered a survey to approximately 2,000 street vendors
while distributing coronavirus relief aid in 2021. The aid was a
one-time payment of \$1,000, available to any individual that owned,
operated, or was otherwise employed by a street vending business in New
York City between 2020 and 2021. First Amendment vendors were also
eligible. There were no restrictions based on residency, the size of the
business, the number of sales, or whether licensed and/or permitted. All
individuals eligible for aid were invited to complete a survey.

The population of street vendors estimated in this paper is the
population eligible for aid (as determined by SVP) that self-identifies
as either a food or merchandise vendor. SVP found aid-eligible
individuals through membership lists, referrals, and canvassing
operations in which SVP affiliates visited neighborhoods. Survey
operations continued until \$2,415,000 in funds were exhausted, yielding
2,060 responses.

The survey included 100 questions and was conducted in nine different
languages: Arabic, Bangla, Cantonese, English, French, Mandarin,
Spanish, Tibetan, and Wolof. The survey items solicited a variety of
information from vendors, such as logistical information (e.g., vending
location, residential location, and frequency of operation), economic
information (e.g., items sold, income, and expenses), and demographic
information (e.g., age, race, ethnicity, and immigration status). Most
relevant is the fact that respondents classified themselves by the goods
they sold (e.g., food vendors, merchandise vendors, First Amendment
vendors, etc.), and respondents indicated whether they had the relevant
licenses and permits to vend. Of the 2,060 responses, 1,400 identified
as food vendors and 559 as merchandise vendors. The remaining 101
respondents were predominantly First Amendment Vendors, which we exclude
from our analysis.

Of the 1,400 food vendors, 349 (25\%) indicated they had a permit to
vend. Of the 559 merchandise vendors, 505 were not veterans, of which
308 (61\%) indicated they had a license to vend. The number of
respondents is listed by neighborhood in Table 1. Neighborhoods with few
respondents are grouped together. A map of the number of respondents by
ZIP Code Tabulation Area (ZIP Code) is shown in the top left panel of
Figure 1. Approximately five percent of vendors did not identify their
vending location. These vendors are included in the New York City total,
but they are excluded from the neighborhood estimates in Table 1 and
Figure 1. For this reason, the Respondent and Population columns do not
sum exactly to the New York City total. Veteran merchandise vendors are
excluded entirely from Table 1 and Figure 1, although these vendors are
reflected in our overall estimate of 23,000 vendors.

\subsection{3. Methodology}\label{methodology}

We estimate the number of street vendors that operate in New York City.
We use ratio estimation, which leverages the fact that the number of
licenses and permits are limited by local law. In Section 3.1, we
provide a simple explanation of ratio estimation. The explanation is
intended to be accessible to a general readership and highlights our
main assumption. In Section 3.2, we develop a theoretical justification
based on the theory of point processes, which while more technical,
allows us to explicitly state the estimands and underlying assumptions
and to derive the standard errors. The details are outlined in Sections
7.1, 7.2, and 7.3.

\subsubsection{3.1 Ratio Estimation}\label{ratio-estimation}

Ratio estimation is a common approach for estimating the size of a
population. See Cochran (1978) and Hald (1998, chap. 16) for a
historical discussion and Lohr (2021, chap. 4) for an introduction. We
provide a simple explanation based on cross-multiplication, also called
the rule of three. The purpose is to highlight the main assumption.

Consider a fixed region \(A\), and let \(N_i(A)\) denote the number of
individuals in region \(A\) of status \(i\). To fix ideas, let \(A\) be
New York City, let \(i = 0\) indicate mobile food vendors without a
permit, and let \(i = 1\) indicate mobile food vendors with a permit.
Also, let \(n_i(A)\) denote the number of respondents of status \(i\).

We observe \(N_1(A)\), \(n_1(A)\), and \(n_0(A)\) from which we would
like to determine \(N_0(A)\), the number of vendors without a permit.
The total number of mobile food vendors with or without a permit is then
\(N_0(A) + N_1(A)\).

We define the ratio estimator as
\[ \hat \Lambda_0(A) = \dfrac{N_1(A) \, n_0(A)}{n_1(A)} . \] It is
called the ratio estimator because it depends on the ratio of random
variables \(n_0(A) \, / \, n_1(A)\). The total number of permitted
vendors, \(N_1(A)\), is considered fixed. We formally define the
estimand that \(\hat \Lambda_0(A)\) estimates in Section 3.2. For now,
we note that the estimator is approximately equal to \(N_0(A)\) when the
response rate, \(p\), is approximately the same for each status, i.e.,
\[ p \approx \frac{n_0(A)}{N_0(A)} \approx \frac{n_1(A)}{N_1(A)} . \]
The formula for \(\hat \Lambda_0(A)\) is obtained by using
cross-multiplication to solve this equation for \(N_0(A)\) in terms of
\(N_1(A)\), \(n_1(A)\), and \(n_0(A)\). It follows that
\(\hat \tau(A) = \hat \Lambda_0(A) + N_1(A)\) is approximately equal to
the total \(N_0(A) + N_1(A)\).

The main assumption of ratio estimation is that the survey data are
representative in the sense that the response rate does not depend on
whether a mobile food vendor has a permit. If we further assume that the
response rate does not vary by subregion, we can also determine the
total number of mobile food vendors by subregion. For example, suppose
the response rate in subregion \(B \subseteq A\) is also approximately
equal to \(p\), i.e.,
\[ p \approx \frac{n_1(A)}{N_1(A)} \approx \frac{n_0(B)}{N_0(B)} . \]
Then the subregion estimator
\[ \tilde \Lambda_0(B) = \dfrac{N_1(A) \, n_0(B)}{n_1(A)} , \] is the
solution for \(N_0(B)\) in terms of \(N_1(A)\), \(n_1(A)\), and
\(n_0(B)\).

The subregion estimator is useful when \(N_1(B)\) is not observed and
\(\hat \Lambda_0(B)\) cannot be calculated directly. This is the case
with the survey data described in Section 2.2. We observe \(N_1(A)\),
the number of permits citywide, but not \(N_1(B)\), the number of
permits within subregion \(B\). Recall that the vending locations of the
respondents are recorded and thus \(n_0(B)\) and \(n_1(B)\) are observed
for any \(B \subseteq A\).

The subregion estimator \(\tilde \Lambda_0 (B)\) works by replacing the
term \(N_1(B)\) in the ratio estimator \(\hat \Lambda_0 (B)\) with
\[\dfrac{N_1(A) \, n_1(B)}{n_1(A)}  ,\] which is justified when
\[ p \approx \frac{n_1(A)}{N_1(A)} \approx \frac{n_1(B)}{N_1(B)} . \]
This suggests the subtotal estimator
\[\tilde \tau (B) = \tilde \Lambda_0(B) + \dfrac{N_1(A) \, n_1(B)}{n_1(A)} = \dfrac{N_1(A) \, \left (n_0(B) + n_1(B) \right )}{n_1(A)} \]
is approximately equal to the subregion total \(N_0(B) + N_1(B)\).

\subsubsection{3.2 Model}\label{model}

We provide a point process justification of the ratio estimation
procedure described in Section 3.1. The purpose is to explicitly state
the estimands and underlying assumptions and to derive the standard
errors.

Let \(\Pi_i\) denote an inhomogeneous Poisson process referencing the
location \(x \in A \subset \mathbb{R}^2\) of each vendor with status
\(i\) at the time the survey was conducted. As in Section 3.1, we fix
ideas by letting \(i = 0\) indicate mobile food vendors without a permit
and \(i = 1\) indicate mobile food vendors with a permit.

The Poisson process assumption may be justified by the law of rare
events. Any sufficiently large area can be partitioned into a large
number of theoretically vendable locations. Whether a vendor is located
within a partition has a vanishingly small probability such that the
number of vendors in that area is well approximated by a Poisson
distribution. This approximation is accurate even under weak dependence
(Freedman 1974). Other justifications are possible. See Section 4.2 for
a justification of the Poisson process based on the theory of birth
processes.

Let \(\lambda_i(x)\) denote the intensity of the process \(\Pi_i\) at
location \(x\) so that the number of individuals in any subregion
\(B \subseteq A\), \(N_i(B)\), is distributed Poisson with mean measure
\(\Lambda_i(B) = \int_B \, \lambda_i(x) \, dx\), i.e.,
\[N_i(B) \sim \text{Poisson} \left ( \int_B \, \lambda_i(x) \, dx \right ) . \]
We assume each individual responds independently to the survey with
probability \(p_i(x) > 0\) such that by Campbell's Theorem (Kingman
1992), the respondents constitute a ``thinned'' inhomogeneous Poisson
process, \(\pi_i\), with intensity \(\lambda_i(x) \, p_i(x)\) at
location \(x\). The number of respondents, \(n_i(B)\), is then
distributed Poisson with mean
\(\int_B \, \lambda_i(x) \, p_i(x) \, dx\), i.e.,
\[n_i(B) \sim \text{Poisson} \left ( \int_B \, \lambda_i(x) \, p_i(x) \, dx \right ) . \]

The data are the locations of the respondents by status (i.e., the
realizations of \(\pi_i\)) and the number of vendors in region \(A\)
with a permit, \(N_1(A)\). In other words, we assume \(n_0(B)\) and
\(n_1(B)\) are observed for any \(B \subseteq A\), while \(N_0(B)\) and
\(N_1(B)\) are unobserved with the exception of \(N_1(A)\).

Our goal is to estimate
\(\mathbb{E}\left[N_0(B) \, | \, N_1(A) \right]\), the expected number
of street vendors in subregion \(B \subseteq A\) without a permit, as
well as
\(\tau(B) = \mathbb{E}\left[N_0(B) + N_1(B) \, | \, N_1(A) \right ]\),
the expected total with or without a permit. We always condition on
\(N_1(A)\) to reflect the fact that while the number of vendors of
either status in any subregion \(B \subsetneq A\) is random, the total
number of vendors with permits citywide is fixed by law. However, we
sometimes suppress the fact that \(N_1(A)\) is fixed in our notation.
For example, \(N_0(B)\) is assumed to be independent of \(N_1(A)\) so we
often write
\(\mathbb{E}\left[N_0(B) \, | \, N_1(A) \right] = \Lambda_0(B)\) and
\(\tau(B) = \Lambda_0(B) + \mathbb{E}\left[N_1(B) \, | \, N_1(A) \right ]\).

In Section 3.1, we described the main assumption of ratio estimation:
the status and spatial distribution of the respondents is
representative. We now make this statement precise. We assume that
\(p_i(x)\) can be decomposed into a constant that does not depend on
\(i\) or \(x\) plus a spatially varying error term,
\(p_i(x) = p \, + \, \epsilon_i(x)\) and that the error term is
orthogonal to the corresponding intensity. i.e.,
\[ \langle \lambda_i, \epsilon_i \rangle = \int_B \lambda_i(x) \, \epsilon_i(x) \, dx = 0 . \]

It follows that
\[n_i(B) \sim \text{Poisson} \left (p \int_B \lambda_i(x) \, dx \right ) \]
so that by conditioning on \(N_1(A)\), we arrive at the following
probability model for the number of respondents with and without permits
\begin{align*}
n_0(B) \sim & \ \text{Poisson} \big (p \, \Lambda_0(B) \big ) \\
n_1(B) \, | \, N_1(A) \sim & \ \text{Binomial} \big (N_1(A), \, p \, q(B)  \big)
\end{align*} where \[ q(B) = \frac{\Lambda_1(B)}{\Lambda_1(A)} .\]

The maximum likelihood estimates for \(\Lambda_0(A)\) and \(p\) are \[
\hat\Lambda_0(A) = \dfrac{N_1(A) \, n_0(A)}{n_1(A)}  \qquad \text{and} \qquad  \hat p = \dfrac{n_1(A)}{N_1(A)} .\]
The estimator \(\hat \Lambda_0(A)\) is asymptotically normal with mean
\(\Lambda_0(A)\) and standard error
\[ \hat{\text{SE}} [ \hat \Lambda_0 (A) ] = \dfrac{N_1(A) \, n_0(A)}{n_1(A)} \sqrt{\dfrac{1}{n_0(A)} + \dfrac{1}{n_1(A)} - \dfrac{1}{N_1(A)}} . \]
The maximum likelihood estimate for the total \(\tau(A)\) is
\(\hat \tau(A) = \hat \Lambda_0(A) + N_1(A)\), which is also
asymptotically normal with mean \(\tau(A)\) and the same standard error
as \(\hat \Lambda_0(A)\). See Section 7.2 for details.

The maximum likelihood estimate for \(\Lambda_0 (B)\) is
\[ \tilde \Lambda_0 (B) = \dfrac{ N_1(A) \, n_0(B)}{n_1(A)} , \] which
is asymptotically normal with mean \(\Lambda_0(B)\) and standard error
\[\tilde{\text{SE}}[\tilde \Lambda_0(B)] = \dfrac{N_1(A) \, n_0(B)}{n_1(A)} \sqrt{ \dfrac{1}{n_0(B)} + \dfrac{1}{n_1(A)} - \dfrac{1}{N_1(A)}} . \]

Finally, the maximum likelihood estimate for the subregion total
\(\tau(B)\) is
\[ \tilde \tau(B) = \dfrac{N_1(A) \, \left ( n_0(B) + n_1(B) \right)}{n_1(A)} , \]
which is also asymptotically normal with mean \(\tau(B)\) and standard
error \[
\tilde{\text{SE}}[\tilde \tau(B)] = \dfrac{N_1(A) \, n_0(B)}{n_1(A)} \sqrt{ \dfrac{1}{n_0(B)} + \frac{1}{n_1(A)} - \frac{1}{N_1(A)} + \dfrac{n_1(B) \, n_1(A \backslash B)}{n_1(A) \, n_0(B)^2}} \]
where \(A \backslash B\) denotes the complement of subregion \(B\). See
Section 7.3 for details.

Note that while we derive asymptotic results for the expected number of
vendors, a prediction interval for \(N_0(B)\) or \(N_0(B) + N_1(B)\)
given \(N_1(A)\) can also be constructed by noting that the distribution
of \(N_i(B)\) is approximately Poisson with rate \(\hat \Lambda_i(B)\).

\subsection{4. Model Validation}\label{model-validation}

We make two assumptions in Section 3. The main assumption is that the
status and spatial distribution of the respondents is representative.
This assumption is sufficient to ensure the ratio estimator is
consistent. A secondary assumption is that the spatial distribution of
vendors is well approximated by a family of inhomogeneous Poisson
processes. This assumption is sufficient to derive the standard errors
analytically.

In this section, we relax both assumptions. Since neither can be relaxed
from the data alone, we consider additional information. In Section 4.1,
we examine a generalization of the ratio estimator in which sampling
weights adjust for differential nonresponse. In Section 4.2, we model
the spatial distribution of respondents using an inhomogeneous negative
binomial process, and we derive a more general formula for the standard
errors in which extra-Poisson variation or ``overdispersion'' can be
explained by the number of vendors that cluster within markets. The
details are outlined in Sections 7.4 and 7.5.

\subsubsection{4.1 Representativeness}\label{representativeness}

The main assumption in Section 3 is that the status and spatial
distribution of the respondents is representative. If this assumption
holds, then the ratio estimator is consistent for a wider class of point
processes in which Campbell's Theorem applies, not just the Poisson
process. See Daley and Vere-Jones (2007) for a general statement of
Campbell's Theorem. If this assumption is violated, then
\(\hat \Lambda_0 (B)\) may be inconsistent. For example, if respondents
are more likely to come from locations in which a higher proportion of
vendors have permits, then the ratio \(n_0(B) \, / \, n_1(B)\) would not
be representative and \(\hat \Lambda_0 (B)\) would likely underestimate
\(\Lambda_0 (B)\), even in large samples.

We think the representativeness assumption is reasonable because the
Street Vendor Project canvassed New York City to distribute a large
amount of relief aid as discussed in Section 2. Nevertheless, to
evaluate how the estimates might change when this assumption is
violated, we generalize the estimator in Section 3.2 to accommodate
spatially varying sampling weights. The generalization allows us to
determine the sensitivity of our estimates to differential nonresponse
as we demonstrate in Section 5.1.

Let \(w_0(x)\) and \(w_1(x)\) denote two spatially varying weight
functions. We define the weighted counts
\[n^w_0(B) = \sum_{x \, \in \, \pi_0 \cap B} w_0(x) \qquad \text{and} \qquad n^w_1(B) = \sum_{x \, \in \, \pi_1 \cap B} w_1(x) \]
where \(\pi_i \cap B\) refers to the points of the process \(\pi_i\)
that fall in subregion \(B \subseteq A\). We also define the weighted
estimators for \(\Lambda_0(A)\) and \(p\),
\[\hat \Lambda_0^w(A) = \dfrac{N_1(A) \, n^w_0(A)}{n^w_1(A)} \qquad \text{and} \qquad \hat p^w = \frac{n^w_1(A)}{N_1(A)} . \]

Under Campbell's Theorem, the expected values \(\mathbb{E}[n^w_0(A)]\)
and \(\mathbb{E}[n^w_1(A) \, | \, N_1(A)]\) are
\[ \mathbb{E} \left [ n^w_0(A)\right ] = \int_{A} \mathbb{E}\bigl[w_0(x)\bigr]\,p_0(x)\,\lambda_0(x)\,dx \qquad \text{and} \qquad \mathbb{E} \left [ n^w_1(A) \, | \, N_1(A) \right ] = \frac{N_1(A)}{\Lambda_1(A)} \, \int_{A} \mathbb{E}\bigl[w_1(x)\bigr]\,p_1(x)\,\lambda_1(x)\,dx .\]

We assume the weight functions \(w_0\) and \(w_1\) can be chosen in such
a way that the expected values are inversely proportional to the
probability of a response. That is,
\[ p = \mathbb{E} [w_0(x) ] \, p_0(x) = \mathbb{E} [w_1(x)] \, p_1(x) . \]
This choice is similar in spirit to Horvitz and Thompson (1952) in that
we seek to correct for the influence of \(p_0(x)\) and \(p_1(x)\). Note
that we only require the average weight to be inversely proportional to
the probability of a response.

It follows that \(\mathbb{E}[n^w_0(A)]\) and
\(\mathbb{E}[n^w_1(A) \, | \, N_1(A)]\) simplify to
\[ \mathbb{E} \left [ n^w_0(A)\right ] = p \, \Lambda_0(A) \qquad \text{and} \qquad \mathbb{E} \left [ n^w_1(A) \, | \, N_1(A) \right ] = p \, N_1(A)\]
and that \(\hat \Lambda_0^w(A)\) is asymptotically normal with mean
\(\Lambda_0(A)\) and standard error
\[\hat{\text{SE}}[\hat \Lambda_0^w(A)] = \dfrac{N_1(A) \, n_0^w(A)}{n_1^w(A)} \sqrt{ \dfrac{\hat \mu^2_0(A) + \hat \sigma_{00}(A,A)}{n_0^w(A)^2} + \dfrac{\hat \mu_1^2(A)}{n^w_1(A)^2} + \dfrac{(N_1(A) - 1) \, \hat \sigma_{11}(A,A)}{N_1(A) \, n^w_1(A)^2}  - \dfrac{1}{N_1(A)} - \dfrac{2 \, \hat \sigma_{01}(A,A)}{n_0^w(A) \, n_1^w(A)} } \]
where
\[ \hat \mu_k^2(A) = \sum_{x \, \in \, \pi_k \cap A} \mathbb{E}\bigl[w_k^2(x)\bigr]  \qquad \text{and} \qquad  \hat \sigma_{kl}(A,A) = \sum_{\substack{x \, \in \, \pi_k \cap A \\[.2em] y \, \in \, \pi_l \cap A \\[.2em] x \neq y }}  \mathbb{C}\bigl(w_k(x),\, w_l(y)\bigr) . \]
See Section 7.4 for details including derivations for the subregion and
subtotal estimators.

The standard error derived in this section for \(\hat \Lambda_0 (A)\) is
identical to the standard error \(\hat \Lambda_0 (A)\) from Section 3.2
if \(w_0(x) = w_1(x) = 1\). In this case,
\(\hat \mu_0^2(A) = n_0(A) = n_0^w(A)\),
\(\hat \mu_1^2(A) = n_1(A) = n^w_1(A)\), and
\(\hat \sigma_{00} = \hat \sigma_{01} = \hat \sigma_{11} = 0\).

Note that \(\hat \Lambda_0(A)\) and \(\hat \Lambda_0^w(A)\) are related
by the relationship
\[\hat \Lambda_0^w(A) = \frac{N_1(A) \ n^w_0(A)}{n^w_1(A)} = \hat \Lambda_0(A) \, \frac{n^w_0(A) \, / \, n_0(A)}{n^w_1(A) \, / \, n_1(A)} .\]
It follows that the relative bias depends only on the ratio of the
average weights \(n^w_0(A) \, / \, n_0(A)\) and
\(n^w_1(A) \, / \, n_1(A)\). For example, if the average permitted
vendor is twice as likely to respond as the average unpermitted vendors,
\(n^w_0(A) \, / \, n_0(A)\) is twice the size of
\(n^w_1(A) \, / \, n_1(A)\) and \(\hat \Lambda_0(A)\) is twice the size
of \(\hat \Lambda_0^w(A)\).

The estimators \(\hat \tau(A)\) and \(\hat \tau^w(A)\) are similarly
related by the relationship
\[\hat \tau^w(A) = \frac{N_1(A) \left ( n^w_0(A) + n^w_1(A) \right )}{n^w_1(A)} = \hat \tau(A) \, \frac{\left ( n^w_0(A) + n^w_1(A) \right ) \, / \, \left ( n_0(A) + n_1(A) \right )}{n^w_1(A) \, / \, n_1(A)} .\]
The same is true for the subregion and subtotal estimators, see Section
7.4.

\subsubsection{4.2 Overdispersion}\label{overdispersion}

The second assumption in Section 3.2 is that the spatial distribution of
vendors is well approximated by a family of inhomogeneous Poisson
processes. We consider this assumption secondary because
\(\hat \Lambda_0 (A)\) is consistent under the wide class of point
processes described in Section 4.1. That said, the standard errors
derived in Section 3.2 do not necessarily hold when the Poisson
assumption is violated.

We think the Poisson assumption is reasonable because a wide class of
data generating processes are well-approximated by a Poisson process
under the law of rare events as discussed in Section 3.2. Nevertheless,
to evaluate how the standard errors might change when this assumption is
violated, we consider an alternative justification based on the theory
of birth processes. As in Section 3.2, we only sketch the justification.
See Ross (2014, chap. 6) for an extended discussion of the theory of
birth processes.

This second justification asks how street vendors arrived at their
present locations, and we consider two possible explanations. We first
suppose that vendors serve customers who congregate at fixed locations,
which we call markets. Let \(M(B)\) denote the number of markets in any
subregion \(B \subseteq A \subset \mathbb{R}^2\). We assume markets grow
over time as vendors arrive to serve customers. If vendor arrivals are
well approximated by an exponential distribution with a constant arrival
rate or ``birth rate'', then the number of vendors at the time of the
survey, \(N_i(B)\), would follow a Poisson distribution as assumed in
Section 3.2.

Suppose instead that larger markets grow faster than smaller markets, an
empirical phenomenon known as preferential attachment. Then the time
between vendor arrivals may be better approximated by an exponential
distribution with a linear arrival rate or Yule process. In this case,
the number of vendors at the time of the survey, \(N_i(B)\), would
follow a negative binomial distribution with mean \(\Lambda_i(B)\) and
variance
\(\Lambda_i(B) \bigr ( \Lambda_i(B) - M(B) \bigr ) \, / \, M(B)\). i.e.,
\[ N_i(B) \sim \ \text{Negative Binomial} \left (\Lambda_i(B), \ \Lambda_i(B) \, \frac{\Lambda_i(B) - M(B)}{M(B)} \right ) \]
where we parameterize the negative binomial distribution by its mean and
variance for comparison with Section 3.2. In this case, \(N_i(B)\)
exhibits extra-Poisson variation or ``overdispersion'' when many vendors
cluster within a market and thus the results of Section 3.2 may not
hold. Note that we define the negative binomial as the total number of
trials required to get a predetermined number of successes when sampling
with replacement.

To derive the standard errors, suppose a portion of the markets, \(p\),
were selected for the survey so that
\[ n_i(B) \sim \ \text{Negative Binomial} \left (p \, \Lambda_i(B), \ p \, \Lambda_i(B) \, \frac{\Lambda_i(B) - M(B)}{M(B)} \right ) . \]

The conditional distribution of \(n_1(B)\) given \(N_1(A)\) is then
\[ n_1(B) \, | \, N_1(A) \sim \  \text{Negative Hypergeometric} \bigl (N_1(A) - 1, \ M(A) - 1, \ p \, q(B) \, M(A) \bigl ) . \]

The conditional mean and variance of \(n_1(A)\) given \(N_1(A)\) is \[
\mathbb{E} \left [ n_1(A) \, | \, N_1(A) \right ] = p \, N_1(A) \qquad \text{and} \qquad \mathbb{V} \bigr ( n_1(A) \, | \, N_1(A) \bigr ) = u_1(A) \, p \, (1 - p) \, N_1(A)\]

so that the ratio estimator
\[\hat \Lambda_0(A) = \frac{N_1(A) \ n_0(A)}{n_1(A)}\] is asymptotically
normal with mean \(\Lambda_0(A)\) and standard error
\[\hat{\text{SE}}[\hat \Lambda_0(A)] = \dfrac{N_1(A) \, n_0(A)}{n_1(A)} \sqrt{ \hat{u}_0(A) \left ( \dfrac{1}{n_0(A)} \right ) + u_1(A) \left( \dfrac{1}{n_1(A)} - \dfrac{1}{N_1(A)} \right )} \]
where
\[\hat{u}_0(B) = \cfrac{\dfrac{N_1(A) \ n_0(B)}{n_1(A)} - M(B)}{M(B)} \qquad \text{and} \qquad u_1(A) = \dfrac{N_1(A) - M(A)}{M(A) + 1} . \]
See Section 7.5 for details including derivations for the subregion and
subtotal estimators. Note that our parameterization makes use of the
relationship \(q(B) = M(B) \, / \, M(A)\).

The standard error derived in this section for \(\hat \Lambda_0 (A)\) is
identical to the standard error for \(\hat \Lambda_0 (A)\) from Section
3.2 except for the weights \(u_0\) and \(u_1\). These weights can be
explained by preferential attachment---the linear ``birth'' rate in
which the number of new vendors is proportional to the number of vendors
already in the market. If there are five vendors in each market on
average, then \(u_0\) and \(u_1\) are approximately 4 and the standard
error of the ratio estimator derived in this section is twice the size
of the standard error derived in Section 3.2. The same is true for
subregion and subtotal estimators, see Section 7.5.

\subsection{5. Data Analysis}\label{data-analysis}

We apply the methodology developed in Section 3 and 4 to the data
described in Section 2. In Section 5.1, we use the ratio estimator from
Section 3.1 and 3.2 to estimate the number of street vendors in New York
City. In Section 5.2, we use the extensions from Section 4.1 and 4.2 to
determine the sensitivity of the estimate to the underlying assumptions.
Sensitivity is assessed using additional data from the NYC Office of
Administrative Trials and Hearings (OATH) (accessed 2024-04-03) and the
American Community Survey from the U.S. Census Bureau (accessed
2022-12-22). We also compare our estimates to those obtained using a
hierarchical model.

\subsubsection{5.1 Estimation}\label{estimation}

According to the survey data described in Section 2.2, \(n_1(A) = 349\)
food vendor respondents indicated they had a permit and
\(n_0(A) = 1,051\) indicated they did not. Since there are
\(N_1(A) = 5,100\) permits available, it follows that the number of
vendors without permits is
\(15,350 \approx 5,100 \times 1,051 \, / \, 349\). Combined with the
\(5,100\) permitted vendors, the total number of mobile food vendors is
approximately \(20,500\).

By similar reasoning---i.e., exploiting the fact that the number of
merchandise licenses is limited for non-veterans---the number of
non-veteran merchandise vendors is estimated to be approximately
\(1,400\). When combined with the estimated \(1,000\) veteran
merchandise vendors (see Section 2.1) and \(20,500\) food vendors, the
total number of street vendors is approximately \(23,000\).

Estimates for select subregions are listed in Table 1. We find that
approximately one quarter of street vendors are located in the Manhattan
neighborhoods of Chelsea, Clinton, and Lower Manhattan. Another quarter
is located in West and North Queens. We also report the margin of error
(i.e., two standard errors, half of the width of a 95\% confidence
interval). The margin of error is calculated using the asymptotic
approximations from Section 3.2. The subregions are chosen so that the
estimated number of vendors in each region is sufficiently large that
the asymptotic approximations are expected to be accurate.

We check our estimates by comparing them to a second, independent
assessment conducted by the Street Vendor Project in the Bronx on May 13
and 15, 2022, in which the number of street vendors was documented
through on-the-ground observations. The independent assessment
identified 188 total street vendors near Fordham Road (Fordham Road BID
and Street Vendor Project 2024). According to the survey data, 17
respondents indicated that they operate near Fordham Road, yielding an
estimate of 166 street vendors total with a standard error of 54. We
conclude this estimate is consistent with the 188 street vendors counted
independently by Fordham Road BID and Street Vendor Project (2024).

\newpage
\setstretch{1}
\renewcommand{\TPTnoteSettings}{}
\begin{ThreePartTable}
\begin{TableNotes}
\item This table provides estimates for the number of mobile food vendors and non-veteran general merchandise vendors. The number of veteran general merchandise vendors is approximately 1,000. First amendment vendors and vendors in NYC Parks are not included.
\end{TableNotes}
\begin{longtable}[t]{rrrr}
\caption{\label{tab:table 2}Number of Street Vendors}\\
\toprule
 & Respondents & Population & Margin of Error\\
\midrule
\addlinespace[0.3em]
\multicolumn{4}{l}{\textbf{Bronx}}\\
\hspace{1em}Bronx Park and Fordham & 68 & 757 & 239\\
\hspace{1em}Southeast Bronx & 28 & 385 & 160\\
\hspace{1em}High Bridge and Morrisania & 20 & 233 & 127\\
\hspace{1em}Central Bronx & 9 & 96 & 81\\
\hspace{1em}Other Bronx & 31 & 264 & 137\\
\textbf{\hspace{1em}Total} & \textbf{156} & \textbf{1,735} & \textbf{376}\\
\midrule
\addlinespace[0.3em]
\multicolumn{4}{l}{\textbf{Brooklyn}}\\
\hspace{1em}\hspace{1em}Northwest Brooklyn & 46 & 613 & 204\\
\hspace{1em}\hspace{1em}Bushwick and Williamsburg & 41 & 433 & 175\\
\hspace{1em}\hspace{1em}Sunset Park & 26 & 344 & 153\\
\hspace{1em}\hspace{1em}\hspace{1em}\hspace{1em}\hspace{1em}\hspace{1em}Flatbush & 19 & 242 & 128\\
\hspace{1em}\hspace{1em}\hspace{1em}\hspace{1em}Borough Park & 10 & 134 & 94\\
\hspace{1em}\hspace{1em}\hspace{1em}\hspace{1em}Southwest Brooklyn & 6 & 64 & 66\\
\hspace{1em}\hspace{1em}\hspace{1em}\hspace{1em}Other Brooklyn & 28 & 374 & 159\\
\hspace{1em}\hspace{1em}\hspace{1em}\hspace{1em}\textbf{Total} & \textbf{176} & \textbf{2,205} & \textbf{418}\\
\midrule\\
\addlinespace[-1em]
\multicolumn{4}{l}{\textbf{Manhattan}}\\
\hspace{1em}\hspace{1em}\hspace{1em}\hspace{1em}Chelsea and Clinton & 341 & 3,289 & 490\\
\hspace{1em}\hspace{1em}\hspace{1em}Lower Manhattan & 105 & 1,250 & 287\\
\hspace{1em}\hspace{1em}\hspace{1em}Gramercy Park and Murray Hill & 97 & 1,062 & 274\\
\hspace{1em}\hspace{1em}\hspace{1em}Lower East Side & 65 & 808 & 232\\
\hspace{1em}\hspace{1em}\hspace{1em}Greenwich Village and Soho & 61 & 726 & 222\\
\hspace{1em}\hspace{1em}\hspace{1em}Upper East Side & 56 & 534 & 194\\
\hspace{1em}\hspace{1em}\hspace{1em}Central Harlem & 46 & 483 & 185\\
\hspace{1em}\hspace{1em}\hspace{1em}Inwood and Washington Heights & 33 & 459 & 176\\
\hspace{1em}\hspace{1em}\hspace{1em}Upper West Side & 48 & 429 & 172\\
\hspace{1em}\hspace{1em}\hspace{1em}East Harlem & 38 & 425 & 174\\
\hspace{1em}\hspace{1em}\hspace{1em}\textbf{Total} & \textbf{890} & \textbf{9,464} & \textbf{879}\\
\midrule\\
\addlinespace[-1em]
\multicolumn{4}{l}{\textbf{Queens}}\\
\hspace{1em}\hspace{1em}\hspace{1em}West Queens & 396 & 4,650 & 692\\
\hspace{1em}\hspace{1em}North Queens & 67 & 896 & 251\\
\hspace{1em}\hspace{1em}Northwest Queens & 23 & 312 & 144\\
\hspace{1em}\hspace{1em}Jamaica & 14 & 181 & 110\\
\hspace{1em}\hspace{1em}Other Queens & 31 & 406 & 168\\
\hspace{1em}\hspace{1em}\textbf{Total} & \textbf{531} & \textbf{6,445} & \textbf{860}\\
\midrule\\
\addlinespace[-1em]
\multicolumn{4}{l}{\textbf{Staten Island}}\\
\hspace{1em}\textbf{\hspace{1em}Total} & \textbf{4} & \textbf{58} & \textbf{58}\\
\midrule
\addlinespace[0.3em]
\multicolumn{4}{l}{\textbf{New York City}}\\
\textbf{\hspace{1em}Total} & \textbf{1,905} & \textbf{21,857} & \textbf{1,941}\\
\bottomrule
\insertTableNotes
\end{longtable}
\end{ThreePartTable}

\newpage

\centering Figure 1: Location of Street Vendors

\vspace{.2 cm}

\includegraphics[width=225px]{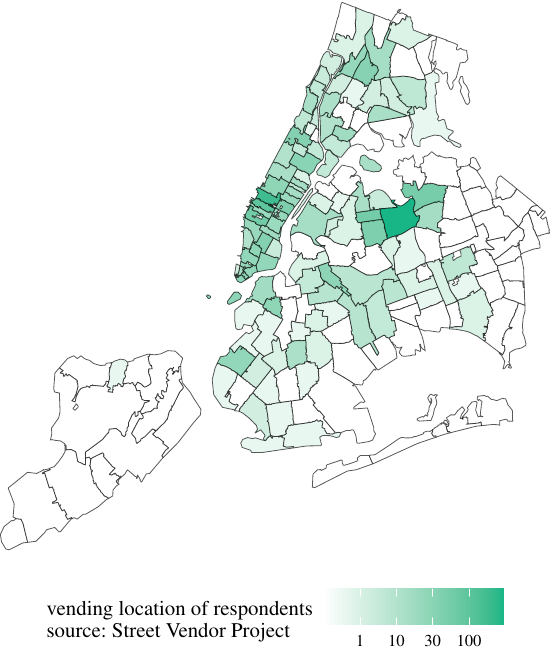}
\includegraphics[width=225px]{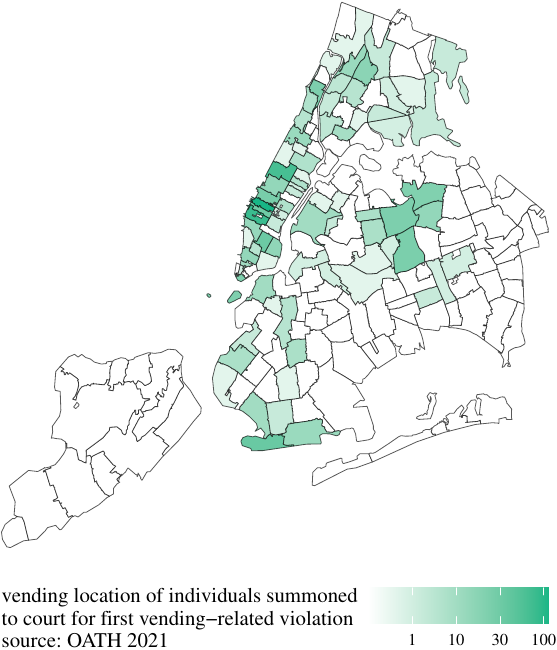}

\vspace{.5 cm}

\includegraphics[width=225px]{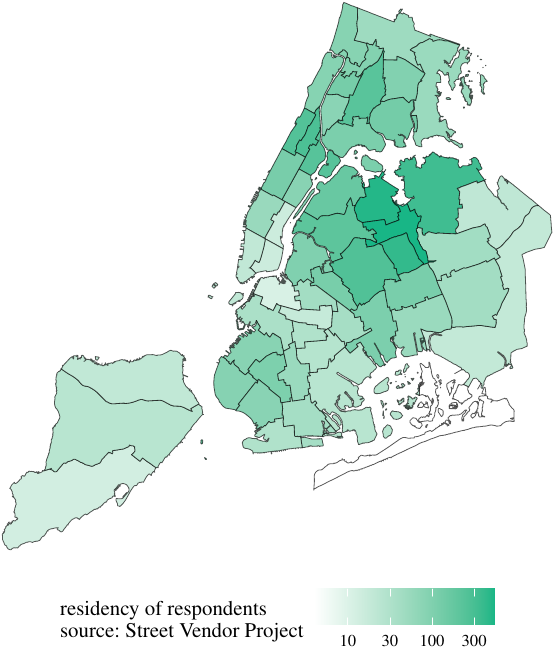}
\includegraphics[width=225px]{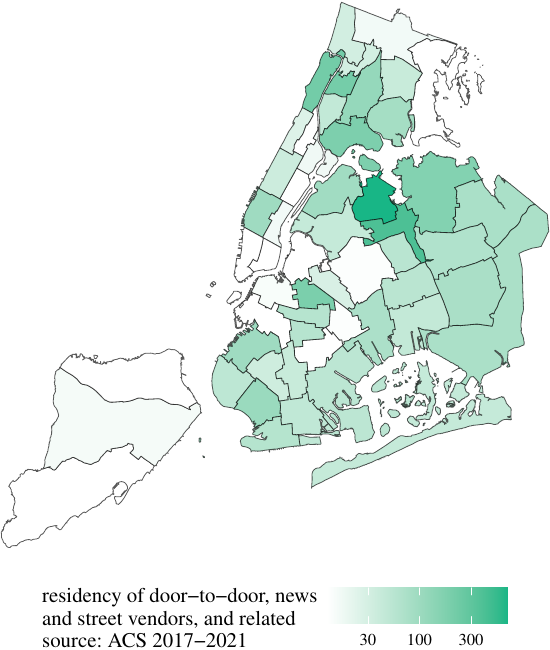}

\begin{quote}
\begin{flushleft}
These maps compare the spatial distribution of respondents (left) to the distribution of vendors suggested by administrative data (New York City Office of Administrative Trials and Hearings (OATH), top right) and federal data (American Community Survey (ACS), bottom right). The top two maps show the log number of vendors by the ZIP Code of their vending location, and the bottom two maps show the log number of vendors by the PUMA (Public Use Microdata Area) of their residency. The top two maps show a similar pattern in which vendors concentrate in North and West Queens and in the Chelsea and Clinton neighborhoods of Manhattan. The bottom two maps show a similar pattern in which residencies are more uniformly distributed in North and West Queens, the South Bronx, and Upper Manhattan.
\end{flushleft}
\end{quote}

\flushleft

\setstretch{1.5}

\subsubsection{5.2 Model Validation}\label{model-validation-1}

We make two assumptions in Section 3. The main assumption is that the
status and spatial distribution of the respondents is representative. We
believe trust in SVP and its mission, along with the aid, removed many
barriers that typically prevent survey operations from reaching
hard-to-reach populations. Instead, nonresponse reflects chance
variation in the personal circumstances of a vendor that are largely
unrelated to location.

Nevertheless, it is possible that canvassing operations systematically
missed locations. We thus compare the spatial distribution of vending
locations reported by respondents to the spatial distribution of vendors
who allegedly violated street vending laws during the year 2021. We
consider the location of an individual's first violation in 2021,
reported in administrative data from the NYC Office of Administrative
Trials and Hearings (OATH) (accessed 2024-04-03). We also compare the
residencies reported by the respondents to the residencies of
Door-To-Door Sales Workers, News And Street Vendors, And Related Workers
(Standard Occupational Classification 41-9091) estimated in the
2017-2021 American Community Survey (ACS) Public Use Micro Sample from
the U.S. Census Bureau (accessed 2022-12-22). These comparisons are
visualized in Figure 1. We find the spatial distribution of vending
locations largely agree (top panels) as do the spatial distribution of
the residencies (bottom panels) in that discrepancies are relatively
minor and explainable by chance. We conclude that if there were a region
of street vendors missed by the SVP, it was also missed by law
enforcement and the ACS.

We can also combine the data from OATH and ACS with the results from
Section 4.1 to assess the sensitivity of the estimates in Section 5.1 to
differential nonresponse as we now demonstrate. While the difference
between these new estimates and the estimates provided in Section 5.1
are not statistically significant at the 95\% level, we believe they
provide insight into how the estimates in Section 5.1 might
theoretically err.

First, we consider the scenario in which the probability a vendor
responds is inversely proportional to the average number of vendors
cited for violating street vending laws in the surrounding ZIP Code each
day. This might be the case if vendors in high enforcement areas are
reluctant to interact with survey takers. The weights are the inverse
probabilities---the average number of enforcement actions taken against
vendors in the surrounding ZIP Code. We find the weighted counts for
food vendors are \(n^w_0(A) = 25.7\) and \(n^w_1(A) = 5.24\). This
suggests there are
\[ \dfrac{\left ( n_0^w(A) + n_1^w(A) \right ) \, / \, \left ( n_0(A) + n_1(A) \right )}{ n_1^w(A) \, / \, n_1(A)} = 47\% \]
more food vendors than estimated in Section 5.1. A similar argument for
merchandise vendors suggests there are 9\% more merchandise vendors than
estimated in Section 5.1. Put together, the weights suggest the number
of street vendors in New York City is 40\% higher than estimated in
Section 5.2.

The second scenario we consider is one in which the probability a vendor
responds is proportional to the average number of vendors cited for
violating street vending laws in the surrounding ZIP Code. This might be
the case if vendors in high enforcement areas are easier to identify and
thus easier to survey. In this scenario, there are 8\% fewer food
vendors than estimated in Section 5.1 and 9\% fewer merchandise vendors,
suggesting the number of street vendors in New York City could be 8\%
lower than estimated in Section 5.2.

The final two scenarios are identical to the first two, except we use
the number of Door-To-Door Sales Workers, News And Street Vendors, And
Related Workers in the surrounding Public Use Microdata Area as
estimated from the 2017-2021 U.S. Census Bureau (accessed 2022-12-22)
(as indicated by the POWPUMA variable). The result is less extreme than
using the average number of vendors cited. If the probability of
response is inversely proportional, there are 33\% more vendors in New
York City than estimated in Section 5.2. If the probability of response
is proportional, there are 5\% fewer vendors. Combining these four
scenarios suggests a conservative range of between 21,000 and 32,200
street vendors operate in New York City. The lower and upper limit of
this range are estimates. Using the standard deviations in Section 4.1,
a conservative 95\% confidence interval is 17,200 to 41,800.

The second assumption is that the spatial distribution of vendors is
well approximated by a family of inhomogeneous Poisson processes. We
argue this assumption is reasonable because a wide class of data
generating processes are well-approximated by a Poisson process under
the law of rare events as discussed in Section 3.2. The fact that many
street vendors operate in a few Zip Codes can be explained by
socioeconomic factors unique to those Zip Codes and does not necessarily
suggest a more complicated model.

Nevertheless, it is possible that vendors cluster within markets
according to a birth process as described in Section 4.2. In that case,
the spatial distribution of vendors may be better approximated by a
family of inhomogeneous negative binomial processes where the weights
\(u_0\) and \(u_1\) correspond to the relative number of vendors without
permits and with permits per market respectively. These weights cannot
be estimated from the survey data alone. However, we believe that if
there is clustering as described in Section 4.2, \(\Lambda_0(A)\) and
\(N_1(A)\) are no more than 10 times larger than \(M(A)\). It follows
that \(u_0\) and \(u_1\) are less than 9, and thus the standard errors
from Section 3.2 and the margins of error from Table 1 are at most 3
times larger. We suspect this bound is conservative, however, and the
typical market contains only one or two vendors.

We conclude this section by comparing our results to a hierarchical
approach in which both the expected number of vendors and the
probability a vendor responds to the survey varies by subregion. The
hierarchical approach provides an alternative framework for constructing
estimates and standard errors, and comparison with the proposed approach
allows us to double check the sensitivity of our estimates and standard
errors to our underlying assumptions. The hierarchical approach still
assumes the independent Poisson processes and their thinned counterparts
as described in Section 3.2. However, we depart from Section 3.2 by
allowing \(p(B_i)\), \(\Lambda_0(B_i)\), and \(\Lambda_1(B_i)\) to vary
according to the model \begin{align*}
\text{logit} \, p(B_i) \sim & \ \text{Normal}(\mu_p, \, \sigma_p^2) \\
\text{log} \, \Lambda_0(B_i)  \sim & \ \text{Normal}(\mu_0, \, \sigma_0^2) \\
\text{log} \, \Lambda_1(B_i)  \sim & \ \text{Normal}(\mu_1, \, \sigma_1^2) .
\end{align*} Note that the hierarchical approach also departs from
Sections 4.1 and 4.2 because \(\Lambda_0(B_i)\) and \(\Lambda_1(B_i)\)
now share a common distribution across subregions so that the
hyperparameters \(\mu_p\) and \(\sigma_p^2\) are identified.

We use the data described in Section 2 and the statistical software
\texttt{Stan} to sample from the posterior distribution of the
hierarchical model. We then use the posterior draws to obtain estimates
and uncertainty intervals. The posterior mean of the total number of
vendors is 21,000 with a 95\% credible interval of 18,900 to 23,900,
which is consistent with the proposed approach. A similar hierarchical
model using the negative binomial process introduced in Section 4.2
produces a similar interval. See Section 7.6 for details including the
\texttt{Stan} code used.

\subsection{6. Discussion}\label{discussion}

We conclude with a discussion. In Section 6.1, we consider how the
theoretical justification provided in Section 3.2 and extended in
Sections 4.1 and 4.2 might be further extended to accommodate more
complex models. In Section 6.2, we consider some implications of the
estimates provided in Section 5.

\subsubsection{6.1 Theoretical Extensions}\label{theoretical-extensions}

In Section 3.2, we develop a theoretical justification for ratio
estimation based on the theory of point processes. We assume the spatial
distribution of survey respondents is representative and
well-approximated by a set of inhomogeneous Poisson processes. Our
approach yields simple, closed-form expressions for the estimates and
standard errors on which policy can be based.

In Section 4, we relax both assumptions to determine the sensitivity of
these expressions. Our relaxations yield additional closed-form
expressions for the estimates and standard errors. These expressions are
more complicated than those derived in Section 3.2. While the estimates
and standard errors are not identified from the survey data alone, they
provide insight into how exactly a violation would change an estimate or
standard error, again facilitating policy.

The results of Section 5 suggest that deviations from our assumptions do
not materially change our findings, except in extreme cases that we do
not believe hold in practice such as if the response probability were
inversely proportional to enforcement. However, it is possible our
findings would change if we were to relax both assumptions further. For
example, in Section 4.1 we relax the assumption that the distribution of
respondents is representative by generalizing the Poisson process to
accommodate probability weights with a known mean and covariance. Future
work might consider the more general case in which the mean and
covariance of the weights are estimated. One might model the response
probability with additional data using logistic regression from which
the mean and covariance of the weights could be estimated. The
additional uncertainty would inflate the standard errors relative to
Section 4.1. See Brick and Montaquila (2009, sec. 3.2) for additional
considerations when modeling the response probability.

For another example, in Section 4.2 we relax the assumption that the
spatial distribution is well-approximated by a set of inhomogeneous
Poisson processes. We consider the possibility that the spatial
distribution of vendors is better approximated by a set of inhomogeneous
negative binomial process, which produces overdispersion when many
vendors cluster within a market. Future work might consider even more
general point process models. For example, one might model the location
of vendors as a Neyman-Scott or Cox process, which also produce
overdispersion. Overdispersion is not inevitable, however, as it is
possible that vendors protect their turf, leading to underdispersion. It
may even be that some vendors benefit from operating near each other,
while others benefit from maintaining their distance. See Cressie and
Wikle (2011, chap. 4.3) for additional considerations when modeling Cox
or dependent processes.

Finally, we assume that spatial regions are sufficiently large or dense
that inference can be achieved using asymptotic arguments. Information
is borrowed across subregions through estimation of a shared ratio.
Future work might compare this approach with other methods for sharing
information between subregions. For example, the literature on small
area estimation often uses hierarchical models such as the one specified
in Section 7.6 to borrow information across subregions, often called
subdomains. See Rao (2003) for for additional considerations when
estimating counts in small regions. One way to augment the hierarchical
approach outlined in Section 5.2 is to model the \(\mu_i\) as a function
of auxiliary data, such as those described in that section.

\subsubsection{6.2 Practical Implications}\label{practical-implications}

We consider the completeness of the New York City Office of
Administrative Trials and Hearings (OATH) and American Community Survey
(ACS) data. Our estimates suggest that while both data sets are
spatially representative, they miss the majority of street vendors. We
provide several explanations for this discrepancy.

We estimate 23,000 vendors work in New York City. In comparison, the
2017-2021 ACS estimates 4,634 individuals work the occupation of
Door-To-Door Sales Workers, News And Street Vendors, And Related Workers
in New York City. This suggests that, according to the ACS, there are at
most 4,634 street vendors in New York City, and thus our finding
indicates that the ACS estimate misses 80\% of street vendors or more.

A fairer comparison might account for the fact that the standard error
of our estimate is approximately 1,000, and the standard error of the
ACS estimate is approximately 488, which we obtained from the replicate
weights. Using the lower limit of a 95\% confidence interval for our
estimate (21,000) and the upper 95\% confidence limit for the ACS
estimate (5,610), our findings suggest the ACS misses roughly three
quarters of street vendors or more.

Note the importance of the assumptions stated in Section 3.2 since
deviations might suggest the ACS misses a smaller or larger portion of
the street vending population. For example, if the average unpermitted
or unlicensed vendor is 10 times more likely to respond than the average
permitted or licensed vendor, then the ACS estimate may miss less than
half of the street vending population. For another example, if the
standard errors from Section 4.2 hold, then the confidence intervals
would overlap if there were more than 300 vendors in the average market.
We believe it is highly unlikely that our assumptions are violated to
this extreme, suggesting the two estimates are inconsistent even under
differential nonresponse and overdispersion.

Additional evidence comes from the OATH data, which indicate
approximately 2,500 (unique) vendors are summoned to court each year for
violating vending laws. If the ACS estimate holds, then more than half
of vendors are summoned to court each year. If our estimate holds, then
approximately one in ten vendors are summoned to court each year. We
believe the latter is more realistic.

There may be several reasons why our estimates are significantly larger
than the ACS. The fact that the Street Vendor Project distributed relief
aid may have encouraged respondents who are unlikely to respond to the
ACS. Moreover, the Street Vendor Project, through membership lists,
referrals, and canvassing may have covered individuals whose households
were missed by the Census Bureau's sampling frame. Indeed, a street
vendor must be sufficiently visible while on the street in order to sell
their items, while the residency of that same vendor may be more easily
overlooked.

Even if the vendor responds to the ACS, they may fail to indicate that
they are a street vendor or work in New York City. Indeed, approximately
25\% of individuals classified as Door-To-Door Sales Workers, News And
Street Vendors, And Related Workers with place of work in New York City
had their occupation imputed due to missing values as indicated by the
occupation allocation flag. Even more, 40\%, had their work location
imputed as indicated by the place of work allocation flag. These
imputation rates are high but not uncommon. Approximately a third of
individuals classified as working in New York City had an occupation
with a higher allocation rate, while 10\% had a higher place of work
allocation rate.

Another explanation is that the Street Vendor Project survey and the ACS
may be measuring different populations or concepts. For example, the ACS
estimate reflects the average number of vendors between 2017 and 2021,
while our estimate reflects 2020 and 2021. For another example, the
Census Bureau determines the occupation of an ACS respondent by
autocoding responses to write-in questions. It is possible that the
Street Vendor Project's definition of a vendor is more inclusive. Street
vending often provides supplemental income, and thus the discrepancy may
reflect the broader challenge of studying the ``gig economy.'' In this
case, either estimate may be relevant, depending on the intended use
case.

\subsection{References}\label{references}

\phantomsection\label{refs}

\subsection{7. Appendix}\label{appendix}

We calculate estimates and standard errors. In Section 7.1, we derive
the ratio estimator as the maximum likelihood estimator. In subsequent
sections, we study the limiting distribution of the ratio estimator and
related statistics as the assumptions underlying these models are
relaxed.

\subsubsection{7.1. Preliminaries}\label{preliminaries}

We consider two independent Poisson processes \(\Pi_0\) and \(\Pi_1\)
and their thinned counterparts \(\pi_0\) and \(\pi_1\) defined on the
region \(A \subset \mathbb{R}^2\) as described in Section 3.2.

Let \(\{B_i\}_{i=1}^{I}\) be an arbitrary partition of \(A\) into
disjoint partition elements such that \(A = \bigcup_{i=1}^{I} B_i\). A
defining characteristic of the Poisson process is that the number of
points in each partition are independent and follow a Poisson
distribution. For example, the counts \(N_1(B_i)\) corresponding with
process \(\Pi_1\) are independent Poisson random variables with mean
\(\Lambda_1(B_i)\), and the counts \(n_1(B_i)\) corresponding with
process \(\pi_1\) are independent Poisson random variables with mean
\(p \, \Lambda_1(B_i)\). Conditioned on the total \(N_1(A)\), the vector
of counts
\(\left [ n_1(B_1), \, \ldots, \, n_1(B_I), \, N_1(A) - n_1(A) \right ]\)
follow a multinomial distribution. See Kingman (1992, sec. 2.4) for
details.

Model 1 is \begin{align*}
n_0(B_i) \sim & \ \text{Poisson} \big (p \, \Lambda_0(B_i) \big ) \\
\left [ n_1(B_1), \, \ldots, \, n_{1}(B_I), \, N_1(A) - n_1(A) \right ] \, | \, N_1(A) \sim & \ \text{Multinomial} \big (N_1(A), \, \left [ p \, q(B_1), \, \ldots, \, p \, q(B_I), \, 1-p \right ]  \big)
\end{align*} where \[ q(B_i) = \frac{ \Lambda_1(B_i)}{\Lambda_1(A)} .\]

The corresponding likelihood \[
\frac{N_1(A)!}{n_1(B_1)! \, \ldots \, n_1(B_I)! \, \left ( N_1(A) - n_1(A) \right )!} \ \left (1-p \right )^{N_1(A) - n_1(A)} \ \prod_{i=1}^I \ \exp(-p \ \Lambda_0(B_i)) \ (p \ \Lambda_0(B_i))^{n_0(B_i)}  \ \frac{1}{n_0(B_i)!} \left (p \, q(B_i) \right )^{n_1(B_i)} \]
with constraint \(\sum_{i=1}^I q(B_i) = 1\) yields maximum likelihood
estimates \begin{equation*}
\tilde \Lambda_0(B_i) = \frac{N_1(A) \ n_0(B_i)}{n_1(A)}, \qquad \hat{p} = \frac{n_1(A)}{N_1(A)}, \qquad \text{and} \qquad \hat{q}(B_i) = \frac{n_1(B_i)}{n_1(A)} . 
\end{equation*}

\subsubsection{7.2 Ratio Estimator}\label{ratio-estimator}

We assume model 1 holds so that the maximum likelihood estimate of
\(\Lambda_0(A) = \sum_{i=1}^I \Lambda_0(B_i)\) is the ratio estimator
\[ \hat \Lambda_0(A) = \sum_{i=1}^I \tilde \Lambda_0(B_i) = \dfrac{N_1(A) \ n_0(A)}{n_1(A)} . \]
Note that \(\hat \Lambda_0(A)\) does not depend on the choice of
partition.

The asymptotic standard error of \(\hat \Lambda_0(A)\) can be obtained
by letting \(\Lambda_0(A)\) and \(N_1(A)\) go to infinity so that \[ 
\begin{bmatrix}
p \, \Lambda_0(A) & 0 \\[2ex]
0 & p \, (1 - p) \, N_1(A)
\end{bmatrix}^{-1/2} 
\begin{bmatrix}
n_0(A) - p \, \Lambda_0 (A)   \\[2ex]
n_1(A) - p \, N_1(A)  
\end{bmatrix} \] converges to a standard bivariate normal distribution.
Note that although \(n_0(A)\) and \(n_1(A)\) are independent, we use
matrix notation to represent the pivotal quantity to be consistent with
later sections.

It follows from the delta method that, \[ \hat \Lambda_0(A) \ 
\dot\sim \ \text{Normal} \left ( \Lambda_0(A), \ \ \begin{bmatrix}
p^{-1} & \ -\dfrac{\Lambda_0(A)}{p \, N_1(A)}
\end{bmatrix}
\begin{bmatrix}
p \, \Lambda_0(A) & 0 \\[3ex] 
0 & p \, (1 - p) \, N_1(A)
\end{bmatrix} 
\begin{bmatrix}
p^{-1} \\[2ex] 
-\dfrac{\Lambda_0(A)}{p \, N_1(A)}
\end{bmatrix} \right ) . \]

The asymptotic standard error simplifies to \[
\text{SE}[\hat \Lambda_0(A)] = \Lambda_0(A) \sqrt{\frac{1}{p \, \Lambda_0(A)} + \frac{1-p}{p \, N_1(A)}} . \]
Substituting \(\hat \Lambda_0(A)\) and \(\hat{p}\) for \(\Lambda_0(A)\)
and \(p\) yields a plug-in estimate of the standard error \[
\hat{\text{SE}}[\hat \Lambda_0(A)] = \dfrac{N_1(A) \, n_0(A)}{n_1(A)} \sqrt{ \dfrac{1}{n_0(A)} + \dfrac{1}{n_1(A)} - \dfrac{1}{N_1(A)}} . \]

\subsubsection{7.3 Subregion Estimator}\label{subregion-estimator}

Let \(B\) denote an arbitrary subregion of \(A\), \(B \subset A\). We
assume that model 1 holds so that the maximum likelihood estimate of
\(\Lambda_0(B)\) is the subregion estimator
\[ \tilde \Lambda_0(B) = \dfrac{N_1(A) \ n_0(B)}{n_1(A)} . \]

The asymptotic standard error of \(\tilde \Lambda_0(B)\) can be obtained
as in Section 7.2 by letting \(\Lambda_0(B)\) and \(N_1(A)\) go to
infinity so that \[ 
\begin{bmatrix}
p \, \Lambda_0(B) & 0 \\[2ex]
0 & p \, (1 - p) \, N_1(A)
\end{bmatrix}^{-1/2} 
\begin{bmatrix}
n_0(B) - p \, \Lambda_0 (B)   \\[2ex]
n_1(A) - p \, N_1(A)  
\end{bmatrix} \] converges to a standard bivariate normal distribution.

It follows from the delta method that \[ \tilde \Lambda_0(B) \ 
\dot\sim \ \text{Normal} \left ( \Lambda_0(B), \ \ \begin{bmatrix}
p^{-1} & \ -\dfrac{\Lambda_0(B)}{p \, N_1(A)}
\end{bmatrix}
\begin{bmatrix}
p \, \Lambda_0(B) & 0 \\[3ex] 
0 & p \, (1 - p) \, N_1(A)
\end{bmatrix} 
\begin{bmatrix}
p^{-1} \\[2ex] 
-\dfrac{\Lambda_0(B)}{p \, N_1(A)}
\end{bmatrix} \right ) . \]

The asymptotic standard error simplifies to \[
\text{SE}[\tilde \Lambda_0(B)] = \Lambda_0(B) \sqrt{\frac{1}{p \, \Lambda_0(B)} + \frac{1-p}{p \, N_1(A)}} . \]
Substituting \(\tilde \Lambda_0(B)\) and \(\hat{p}\) for
\(\Lambda_0(B)\) and \(p\) yields a plug-in estimate of the standard
error \[
\tilde{\text{SE}}[\tilde \Lambda_0(B)] = \dfrac{N_1(A) \, n_0(B)}{n_1(A)} \sqrt{ \dfrac{1}{n_0(B)} + \dfrac{1}{n_1(A)} - \dfrac{1}{N_1(A)}} . \]

The expected total in subregion \(B\) is
\(\Lambda_0 (B) + q(B) \, N_1(A)\). Substituting the maximum likelihood
estimates \(\tilde \Lambda_0(B)\) and
\[\hat q(B) \, N_1(A) = \dfrac{N_1(A) \, n_1(B)}{n_1(A)} \] yields the
estimated total in subregion \(B\),
\[ \tilde \tau(B) =  \dfrac{N_1(A) \, \left ( n_0(B) + n_1(B) \right)}{n_1(A)} =  \dfrac{N_1(A) \, \left ( n_0(B) + n_1(B) \right)}{n_1(B) + n_1(A \backslash B)} \]
where \(A \backslash B\) denotes the complement of subregion \(B\).

As before, the asymptotic standard error of \(\tilde \tau(B)\) can be
obtained by noting that \[ 
\begin{bmatrix}
p \, \Lambda_0(B) & 0 & 0 \\[2ex]
0 & p \, q(B) \, (1 - p \, q(B)) \, N_1(A) & - p^2 \, q(B) \, q(A \backslash B) \, N_1(A) \\[2ex]
0 & - p^2 \, q(B) \, q(A \backslash B) \, N_1(A) & p \, q(A \backslash B) \, (1 - p \, q(A \backslash B)) \, N_1(A) \\[2ex]
\end{bmatrix}^{-1/2} 
\begin{bmatrix}
n_0(B) - p \, \Lambda_0 (B) \\[2ex]
n_1(B) - p \, q(B) \, N_1(A) \\[2ex]
n_1(A \backslash B) - p \, q(A \backslash B) \, N_1(A) \\[2ex]
\end{bmatrix} \] converges to a standard trivariate normal distribution.

It follows from the delta method that \(\tilde \tau (B)\) is
approximately normal with mean
\(\Lambda_0(B) + \mathbb E[N_1(B) \, | \, N_1(A)]\) and variance \[
\begin{aligned}
& \begin{bmatrix}
\displaystyle p^{-1} &
\displaystyle \frac{q(A \backslash B) \, N_1(A) - \Lambda_0(B)}{p \, N_1(A)} &
\displaystyle -\frac{\Lambda_0(B) + q(B) \, N_1(A)}{p \, N_1(A)}
\end{bmatrix}
\\[2ex]
&\qquad \times
\begin{bmatrix}
p \, \Lambda_0(B) & 0 & 0 \\[2ex]
0 & p \, q(B) \, (1 - p \, q(B)) \, N_1(A) & - p^2 \, q(B) \, q(A \backslash B) \, N_1(A) \\[2ex]
0 & - p^2 \, q(B) \, q(A \backslash B) \, N_1(A) & p \, q(A \backslash B) \, (1 - p \, q(A \backslash B)) \, N_1(A)
\end{bmatrix}
\\[2ex]
&\qquad \times
\begin{bmatrix}
p^{-1} \\[2ex] 
\dfrac{q(A \backslash B) \, N_1(A) - \Lambda_0(B)}{p \, N_1(A)} \\[4ex]
-\dfrac{\Lambda_0(B) + q(B) \, N_1(A)}{p \, N_1(A)}
\end{bmatrix}.
\end{aligned}
\]

The asymptotic standard error simplifies to \[
\text{SE}[\tilde \tau(B)] = \Lambda_0(B) \sqrt{\frac{1}{p \, \Lambda_0(B)} + \frac{1 - p}{p \, N_1(A)} + \frac{q(B) \, q(A \backslash B) \, N_1(A)}{p \, \Lambda_0(B)^2}} . \]
Substituting \(\tilde \Lambda_0\), \(\hat{p}\), \(\hat{q}\)
\(\tilde N_1\) for \(\Lambda_0\), \(p\), \(q\), and \(N_1\) yields a
plug-in estimate of the standard error \[
\tilde{\text{SE}}[\tilde \tau(B)] = \dfrac{N_1(A) \, n_0(B)}{n_1(A)} \sqrt{ \dfrac{1}{n_0(B)} + \frac{1}{n_1(A)} - \frac{1}{N_1(A)} + \dfrac{n_1(B) \, n_1(A \backslash B)}{n_1(A) \, n_0(B)^2}} . \]

\subsubsection{7.4 Weighted Estimation}\label{weighted-estimation}

We consider a generalization of model 1 in which the spatial
distribution of respondents is not representative. Let
\(\{B_i\}_{i=1}^{I}\) be an arbitrary partition of \(A\) into disjoint
partition elements such that \(A = \bigcup_{i=1}^{I} B_i\). Model 2 is
\begin{align*}
n_0(B_i) \sim & \ \text{Poisson} \left ( \int_{B_i} \, p_0(x) \, \lambda_0(x) \, dx \right ) \\
\left [ n_1(B_1), \, \ldots, \, n_{1}(B_I), \, N_1(A) - n_1(A) \right ] \, | \, N_1(A) \sim & \ \text{Multinomial} \big (N_1(A), \, \left [ r(B_1), \, \ldots, \, r(B_I), \, 1-r(A) \right ]  \big) 
\end{align*} where
\[r(B_i) = \frac{\int_{B_i} \, p_1(x) \, \lambda_1(x) \, dx}{\Lambda_1(A)} .\]

We define estimators
\[\hat \Lambda_0^w(A) = \dfrac{N_1(A) \, n^w_0(A)}{n^w_1(A)} \qquad \text{and} \qquad \hat{p}^w = \frac{n^w_1(A)}{N_1(A)} \]
where
\[n^w_0(B) = \sum_{x \, \in \, \pi_0 \cap B} w_0(x) \qquad \text{and} \qquad n^w_1(B) = \sum_{x \, \in \, \pi_1 \cap B} w_1(x) . \]
The notation \(\pi_i \cap B\) refers to the points of the process
\(\pi_i\) that fall in subregion \(B \subseteq A\). The weighted counts
\(n^w_0(B)\) and \(n^w_1(B)\) equal the unweighted counts \(n_0(B)\) and
\(n_1(B)\) when \(w_0(x) = w_1(x) = 1\).

We allow the weight functions \(w_0\) and \(w_1\) to be random, but we
assume they are nonnegative, uniformly bounded away from \(0\) and
\(\infty\) in expectation, and independent of \(\pi_0\) and \(\pi_1\).
We also assume the dependence between \(w_{0}(x)\) and both \(w_{0}(y)\)
and \(w_{1}(y)\) is sufficiently weak for \(x \neq y\). For example, the
weights could be stationary and the covariances
\(\mathbb{C}(w_{0}(x), \, w_{0}(y))\) and
\(\mathbb{C}(w_{0}(x), \, w_{1}(y))\) decay at an exponential rate. See
Lehmann (1999, sec. 2.8) for a discussion of these and related
conditions.

Under Campbell's Theorem, the mean and variance of the \(n^w_0(B_i)\)
are
\[ \mathbb{E} \left [ n^w_0(B_i)\right ] = \mu_0(B_i) \qquad \text{and} \qquad \mathbb{V} \bigr ( n^w_0(B_i) \bigr ) =  \mu^2_0(B_i) + \sigma_{00}(B_i,B_i)  . \]

The conditional mean and variance of the \(n^w_1(B_i)\) given \(N_1(A)\)
are
\[ \mathbb{E} \left [ n^w_1(B_i) \, | \, N_1(A) \right ] = \frac{N_1(A)}{\Lambda_1(A)} \, \mu_{1}(B_i) \]
and
\[\mathbb{V} \bigr ( n_1^w(B_i) \, | \, N_1(A) \bigr ) = \frac{N_1(A)}{\Lambda_1(A)} \, \mu^2_1(B_i) - \frac{N_1(A)}{\Lambda_1(A)^2} \, \mu_1(B_i)^2  + \frac{N_1(A) \left (N_1(A) - 1 \right)}{\Lambda_1(A)^2} \, \sigma_{11}(B_i, B_i) .
\]

The conditional covariances between \(n_0^w(B_i)\) and \(n_1^w(B_i)\)
and between \(n_1^w(B_i)\) and \(n_1^w(B_j)\) given \(N_1(A)\) are \[
\mathbb{C} \bigr ( n_0^w(B_i), \, n_1^w(B_i) \, | \, N_1(A) \bigr ) = \frac{N_1(A)}{\Lambda_1(A)} \, \sigma_{01}(B_i, B_i) \]
and
\[\mathbb{C} \bigr ( n_1^w(B_i), \, n_1^w(B_j) \, | \, N_1(A) \bigr ) =  \frac{N_1(A) \left ( N_1(A) - 1 \right )}{\Lambda_1(A)^2} \,\sigma_{11}(B_i, B_j) \]

where \[\mu_k(B_i) 
= \int_{B_i} \mathbb{E}\bigl[w_k(x)\bigr]\,p_k(x)\,\lambda_k(x)\,dx
\qquad
\mu_k^2(B_i) 
= \int_{B_i} \mathbb{E}\bigl[w_k^2(x)\bigr]\,p_k(x)\,\lambda_k(x)\,dx\]
and \[\sigma_{kl}(B_i, B_j) 
= \int_{B_i \times B_j} \mathbb{C}\bigl(w_k(x),\, w_l(y)\bigr)\,p_k(x)\,\lambda_k(x)\,p_l(y)\,\lambda_l(y)\,dx\,dy . \]

Finally, we assume the weight functions \(w_0\) and \(w_1\) are chosen
so that the expected value is inversely proportional with the
probability of a response. That is,
\[ p = \mathbb{E} [w_0(x) ] \, p_0(x) = \mathbb{E} [w_1(x)] \, p_1(x) \]
and thus \(\mu_k(B_i) = p \, \Lambda_k(B_i)\).

The asymptotic distribution of \(\hat \Lambda_0^w(A)\) can now be
obtained by letting \(\Lambda_0(B)\) and \(N_1(A)\) go to infinity so
that, as long as no vanishing subregion dominates, under the Anscombe
central limit theorem for randomly indexed sums (Sugiman 1986),

\[
\begin{bmatrix} 
\mu^2_0(A) + \sigma_{00}(A,A)  & \qquad \dfrac{N_1(A) \, \sigma_{01}(A,A)}{\Lambda_1(A)} \\[7ex]
\dfrac{N_1(A) \, \sigma_{01}(A,A)}{\Lambda_1(A)} & \qquad
\dfrac{N_1(A) \, \mu^2_1(A)}{\Lambda_1(A)} - p^2 \, N_1(A)  + \dfrac{\left ( N_1(A)^2 - N_1(A) \right ) \, \sigma_{11}(A,A)}{\Lambda_1(A)^2} 
\end{bmatrix}^{-1/2} 
\begin{bmatrix}
n_0^w(A) - p \, \Lambda_0(A) \\[9ex]
n_1^w(A) - p \, N_1(A)   
\end{bmatrix}
\] converges to a standard bivariate normal distribution.

It follows from the delta method that \(\hat \Lambda_0^w(A)\) is
approximately normal with mean \(\Lambda_0(A)\) and variance

\[
\begin{bmatrix}
p^{-1} & \ -\dfrac{\Lambda_0(A)}{p \, N_1(A)}
\end{bmatrix}
\begin{bmatrix} 
\mu^2_0(A) + \sigma_{00}(A,A)  & \qquad \dfrac{N_1(A) \, \sigma_{01}(A,A)}{\Lambda_1(A)} \\[7ex]
\dfrac{N_1(A) \, \sigma_{01}(A,A)}{\Lambda_1(A)} & \qquad
\dfrac{N_1(A) \, \mu^2_1(A)}{\Lambda_1(A)} - p^2 \, N_1(A)  + \dfrac{\left ( N_1(A)^2 - N_1(A) \right ) \, \sigma_{11}(A,A)}{\Lambda_1(A)^2} 
\end{bmatrix}
\begin{bmatrix}
p^{-1} \\[8ex] 
-\dfrac{\Lambda_0(A)}{p \, N_1(A)}
\end{bmatrix}.
\]

The asymptotic standard error, \(\text{SE}[\hat \Lambda_0^w(A)]\),
simplifies to
\[\Lambda_0(A) \sqrt{  \dfrac{\mu^2_0(A) + \sigma_{00}(A,A)}{p^2 \, \Lambda_0(A)^2}  +  \dfrac{\mu_1^2(A)}{p^2 \, N_1(A) \, \Lambda_1(A)} + \dfrac{ (N_1(A) - 1) \, \sigma_{11}(A,A)}{p^2 \, N_1(A) \, \Lambda_1(A)^2} - \dfrac{1}{N_1(A)} - \dfrac{2 \, \sigma_{01}(A,A)}{p^2 \, \Lambda_0(A) \, \Lambda_1(A)} } . \]
Substituting \(\hat \Lambda_0^w(A)\), \(\hat{p}^w\), and \(N_1(A)\) for
\(\Lambda_0(A)\), \(p\), and \(\Lambda_1(A)\), yields a plug-in estimate
of the standard error,
\[\hat{\text{SE}}[\hat \Lambda_0^w(A)] = \dfrac{N_1(A) \, n_0^w(A)}{n_1^w(A)} \sqrt{ \dfrac{\hat \mu^2_0(A) + \hat \sigma_{00}(A,A)}{n_0^w(A)^2} + \dfrac{\hat \mu_1^2(A)}{n^w_1(A)^2} + \dfrac{(N_1(A) - 1) \, \hat \sigma_{11}(A,A)}{N_1(A) \, n^w_1(A)^2}  - \dfrac{1}{N_1(A)} - \dfrac{2 \, \hat \sigma_{01}(A,A)}{n_0^w(A) \, n_1^w(A)} } \]
where
\[ \hat \mu_k^2(A_i) = \sum_{x \, \in \, \pi_k \cap A_i} \mathbb{E}\bigl[w_k^2(x)\bigr] 
\qquad \text{and} \qquad  \hat \sigma_{kl}(A_i,A_j) = \sum_{\substack{x \, \in \, \pi_k \cap A_i \\[.2em] y \, \in \, \pi_l \cap A_j \\[.2em] x \neq y }}  \mathbb{C}\bigl(w_k(x),\, w_l(y)\bigr).  \]

For weighted subregion and subtotal estimates, we define additional
estimators
\[\tilde \Lambda_0^w(B) = \dfrac{N_1(A) \, n^w_0(B)}{n^w_1(A)} \qquad \text{and} \qquad \hat{q}^w(B) = \frac{n^w_1(B)}{n^w_1(A)} . \]
By an argument analogous to Section 7.3, the subregion estimator
\(\tilde \Lambda_0^w(B)\) is approximately normal with mean
\(\Lambda_0(B)\) and standard error,
\(\text{SE}[\tilde \Lambda_0^w(B)]\),
\[\Lambda_0(B) \sqrt{  \dfrac{\mu^2_0(B) + \sigma_{00}(B,B)}{p^2 \, \Lambda_0(B)^2}  +  \dfrac{\mu_1^2(A)}{p^2 \, N_1(A) \, \Lambda_1(A)} + \dfrac{(N_1(A) - 1) \, \sigma_{11}(A,A)}{p^2 \, N_1(A) \, \Lambda_1(A)^2} - \dfrac{1}{N_1(A)} - \dfrac{2 \, \sigma_{01}(B,A)}{p^2 \, \Lambda_0(B) \, \Lambda_1(A)} } . \]

Substituting \(\tilde \Lambda_0^w(B)\), \(\hat{p}^w\), and \(N_1(A)\)
for \(\Lambda_0(B)\), \(p\), and \(\Lambda_1(A)\) yields a plug-in
estimate of the standard error
\[\tilde{\text{SE}}[\tilde \Lambda_0^w(B)] = \dfrac{N_1(A) \, n_0^w(B)}{n_1^w(A)} \sqrt{ \dfrac{\hat \mu^2_0(B) + \hat \sigma_{00}(B,B)}{n_0^w(B)^2} + \dfrac{\hat \mu_1^2(A)}{n^w_1(A)^2} + \dfrac{(N_1(A) - 1) \, \hat \sigma_{11}(A,A)}{N_1(A) \, n^w_1(A)^2} - \dfrac{1}{N_1(A)} - \dfrac{2 \, \hat \sigma_{01}(B,A)}{n_0^w(B) \, n_1^w(A)} }. \]

By another argument analogous to Section 7.3, the subregion total,
\[ \tilde \tau^w(B) = \dfrac{N_1(A) \, \left ( n_0^w(B) + n_1^w(B) \right)}{n_1^w(B) + n_1^w(A \backslash B)} , \]
is approximately normal with mean
\(\Lambda_0(B) + \mathbb E[N_1(B) \, | \, N_1(A]\) and standard error
\[\text{SE}[\tilde \tau(B)] = \Lambda_0(B) \sqrt{  \dfrac{\mu^2_0(B) + \sigma_{00}(B,B)}{p^2 \, \Lambda_0(B)^2}  +  \dfrac{v_1 + v_2 + 2 \, N_1(A) \, v_3}{p^2 \, N_1(A) \, \Lambda_0(B)^2 \, \Lambda_1(A)}} \]
where \[ \begin{aligned}
v_1 & =  \left ( q(A \backslash B) \, N_1(A) - \Lambda_0(B) \right)^2 \left ( \mu_1^2(B) - p^2 \, q(B)^2 \, \Lambda_1(A) + \dfrac{(N_1(A) - 1) \, \sigma_{11}(B, B)}{\Lambda_1(A)} \right ) \\[2em]
v_2 & =  \left ( \Lambda_0(B) + q(B) \, N_1(A) \right)^2 \left ( \mu_1^2(A \backslash B) - p^2 \, q(A \backslash B)^2 \, \Lambda_1(A) + \dfrac{(N_1(A) - 1) \, \sigma_{11}(A \backslash B, A \backslash B)}{\Lambda_1(A)} \right ) \\[2em]
v_3 &= \left ( q(A \backslash B) \, N_1(A) - \Lambda_0(B) \right ) \sigma_{01}(B, B) - \left ( \Lambda_0(B) + q(B) \, N_1(A)  \right ) \sigma_{01}(B, A \backslash B) \\[1ex] 
& \qquad - \frac{\left ( q(A \backslash B) \, N_1(A) - \Lambda_0(B) \right ) \left ( \Lambda_0(B) + q(B) \, N_1(A)  \right ) \left (N_1(A) - 1 \right) \sigma_{11}(B, A \backslash B) }{\Lambda_1(A)}. \\[2ex] \end{aligned} \]

Substituting \(\tilde \Lambda_0^w(B)\), \(\hat{p}^w\), \(\hat q^w(B)\),
\(\hat q^w(A \backslash B)\), and \(N_1(A)\) for \(\Lambda_0(B)\),
\(p\), \(q(B)\), \(q(A \backslash B)\), and \(\Lambda_1(A)\) yields a
plug-in estimate of the standard error \[
\tilde{\text{SE}}[\tilde \tau^w(B)] = \dfrac{N_1(A) \, n_0^w(B)}{n_1^w(A)} \sqrt{ \dfrac{\hat \mu^2_0(B) + \hat \sigma_{00}(B,B)}{n_0^w(B)^2} + \dfrac{\hat \mu_1^2(A)}{n^w_1(A)^2} +  \dfrac{\hat v_1 + \hat v_2 + 2 \, N_1(A) \, \hat v_3}{n^w_0(B)^2 \, N_1(A)^2}} \]
where \[ \begin{aligned}
\hat v_1 & =  \left ( n^w_1(A \backslash B) - n^w_0(B) \right)^2 \left ( \dfrac{N_1(A)}{n^w_1(A)} \right )^2 \left (\hat \mu_1^2(B) + \dfrac{(N_1(A) - 1) \, \hat \sigma_{11}(B, B) - n_1^w(B)^2}{N_1(A)} \right ) \\[2em]
\hat v_2 & =  \left ( n^w_0(B) +  n^w_1(B) \right)^2 \left ( \dfrac{N_1(A)}{n^w_1(A)} \right )^2 \left ( \hat \mu_1^2(A \backslash B) + \dfrac{(N_1(A) - 1) \, \hat \sigma_{11}(A \backslash B, A \backslash B) - n_1^w(A \backslash B)^2}{N_1(A)} \right ) \\[2em]
\hat v_3 &=  \dfrac{N_1(A)}{n^w_1(A)} \bigg ( \left ( n^w_1(A \backslash B) - n^w_0(B) \right) \hat \sigma_{01}(B, B) - \left ( n^w_0(B) + n^w_1(A \backslash B) \right) \hat \sigma_{01}(B, A \backslash B) \\[1ex] 
& \qquad \qquad \qquad  - \left ( n^w_1(A \backslash B) - n^w_0(B) \right) \left ( n^w_0(B) + n^w_1(A \backslash B) \right) \frac{\left (N_1(A) - 1 \right) \hat \sigma_{11}(B, A \backslash B) }{n^w_1(A)} \bigg ). \\[2ex] \end{aligned} \]

\subsubsection{7.5 Overdispersion}\label{overdispersion-1}

We consider a generalization of model 1 in which the number of
individuals in each area may exhibit extra-Poisson variation or
``overdispersion.'' Let \(\{B_i\}_{i=1}^{I}\) be an arbitrary partition
of \(A\) into disjoint partition elements such that
\(A = \bigcup_{i=1}^{I} B_i\). Model 3 is \[ \begin{aligned} 
n_0(B_i) \sim & \ \text{Negative Binomial} \bigl (p \, \Lambda_0(B_i), \ u_{0}(B_i) \, p \, \Lambda_0(B_i)  \bigr ) \\ 
\bigl [ n_1(B_1), \, \ldots, \, n_1(B_I), \, N_1(A) - n_1(A) \, \bigl ] \, | \, N_1(A) \sim & \ \text{MNH} \bigr ( N_1(A) - 1, \ \bigl [ p \, q(B_1) \, M(A), \, \ldots, \ p \, q(B_I) \, M(A), \ (1-p) \, M(A) \bigl ] \bigr )
\end{aligned} \] where
\[u_0(B_i) = \frac{\Lambda_0(B_i) - M(B_i)}{M(B_i)}\] and MNH denotes
the multivariate negative hypergeometric distribution. See Johnson,
Kotz, and Balakrishnan (1997, chap. 35, page 81). Note that we define
the negative binomial and negative hypergeometric distributions as the
total number of trials required to get a predetermined number of
successes when sampling with and without replacement, respectively.

The mean and variance of \(n_0(A)\) and \(n_1(A)\) given \(N_1(A)\) are
\[
\begin{aligned}
\mathbb{E}\bigl[n_0(A)\bigr]
&= p\,\Lambda_0(A)
&\qquad
\mathbb{V}\bigl(n_0(A)\bigr)
&= u_0(A)\,p\,\Lambda_0(A)
\\[6pt]
\mathbb{E}\bigl[n_1(A)\mid N_1(A)\bigr]
&= p\,N_1(A)
&\qquad
\mathbb{V}\bigl(n_1(A)\mid N_1(A)\bigr)
&= u_1(A)\,p\,(1-p)\,N_1(A)
\end{aligned}
\] where \[u_1(A) = \dfrac{N_1(A) - M(A)}{M(A) + 1} . \]

The asymptotic standard error of \(\hat \Lambda_0(A)\) can be obtained
as in Section 7.2 by letting \(\Lambda_0(A)\), \(M(A)\), and \(N_1(A)\)
go to infinity so that \[ 
\begin{bmatrix} 
u_0(A) \, p \, \Lambda_0(A) & 0 \\[4ex]
0 & u_1(A) \, p \, (1 - p) \, N_1(A)
\end{bmatrix}^{-1/2} 
\begin{bmatrix}
n_0(A) - p \, \Lambda_0(A)   \\[5ex]
n_1(A) - p \, N_1(A)  
\end{bmatrix} \] converges to a standard bivariate normal distribution.
(The triple \(\left [ n_0(A), \, n_1(A), \, N_1(A) - n_1(A) \right ]\)
is asymptotically normal from which we can deduce the asymptotic
normality of \(\left [ n_0(A), \, n_1(A) \right ] \, | \, N_1(A)\). See
Feller (1968, chap. 7, page 194). Note that we assume
\(N_1(A) \, / \, M(A)\) converges to a finite constant \(c > 1\) so that
the subregion counts \(n_1(B_i)\) are negatively dependent given
\(N_1(A)\), even in the limit.)

It follows from the delta method that \[ \hat \Lambda_0(A) \ 
\dot\sim \ \text{Normal} \left ( \Lambda_0(A), \ \ \begin{bmatrix}
p^{-1} & \ -\dfrac{\Lambda_0(A)}{p \, N_1(A)}
\end{bmatrix}
\begin{bmatrix}
u_0(A) \, p \, \Lambda_0(A)  & 0 \\[3ex] 
0 & u_1(A) \, p \, (1 - p) \, N_1(A)
\end{bmatrix} 
\begin{bmatrix}
p^{-1} \\[2ex] 
-\dfrac{\Lambda_0(A)}{p \, N_1(A)}
\end{bmatrix} \right ) . \] The asymptotic standard error simplifies to
\[\text{SE}[\hat \Lambda_0(A)] = \Lambda_0(A) \sqrt{ u_0(A) \left ( \frac{1}{p \, \Lambda_0(A)} \right ) + u_1(A) \left ( \frac{1-p}{p \, N_1(A)} \right ) } . \]
Substituting \(\hat \Lambda_0(A)\) and \(\hat{p}\) for \(\Lambda_0(A)\)
and \(p\) yields a plug-in estimate of the standard error \[
\hat{\text{SE}}[\hat \Lambda_0(A)] = \dfrac{N_1(A) \, n_0(A)}{n_1(A)} \sqrt{ \hat{u}_0(A) \left ( \dfrac{1}{n_0(A)} \right ) + u_1(A) \left( \dfrac{1}{n_1(A)} - \dfrac{1}{N_1(A)} \right )} \]
where
\[\hat{u}_0(B) = \cfrac{\dfrac{N_1(A) \ n_0(B)}{n_1(A)} - M(B)}{M(B)} . \]

By an argument analogous to Section 7.3, the subregion estimator
\[ \tilde \Lambda_0(B) = \dfrac{N_1(A) \ n_0(B)}{n_1(A)} \] is
approximately normal with mean \(\Lambda_0(B)\) and standard error
\[\text{SE}[\tilde \Lambda_0(B)] = \Lambda_0(B) \sqrt{ u_0(B) \left ( \frac{1}{p \, \Lambda_0(B)} \right ) + u_1(A) \left ( \frac{1-p}{p \, N_1(A)} \right ) } . \]

Substituting \(\tilde \Lambda_0(B)\) and \(\hat{p}\) for
\(\Lambda_0(B)\) and \(p\) yields a plug-in estimate \[
\tilde{\text{SE}}[\tilde \Lambda_0(B)] = \dfrac{N_1(A) \, n_0(B)}{n_1(A)} \sqrt{ \hat{u}_0(B) \left ( \dfrac{1}{n_0(B)} \right ) + u_1(A) \left( \dfrac{1}{n_1(A)} - \dfrac{1}{N_1(A)} \right )} . \]

Finally, we examine the behavior of the subregion total
\[\tilde{\tau}(B) = \frac{N_1(A) \, \left ( n_0(B) + n_1(B) \right )}{n_1(A)} . \]
The mean, variance, and covariance of \(n_1(B)\) and
\(n_1(A \backslash B)\) given \(N_1(A)\) are \begin{align*}
\mathbb{E} \left [ n_1(B) \, | \, N_1(A) \right ] &= p \, q(B) \, N_1(A) \\
\mathbb{V} \bigr ( n_1(B) \, | \, N_1(A) \bigr ) &= u_1(A) \, p \, q(B) \, \left ( 1 - q(B) \, p \right ) \, N_1(A) \bigl ) \\
\mathbb{C} \bigr ( n_1(B), \, n_1(A \backslash B) \, | \, N_1(A) \bigr ) &= -u_1(A) \, p^2 \, q(B) \, q(A \backslash B) \, N_1(A) .
\end{align*}

By another argument analogous to Section 7.3, \(\tilde \tau(B)\) is
approximately normal with mean
\(\Lambda_0(B) + \mathbb E [ N_1(B) \, | \, N_1(A) ]\) and standard
error
\[\text{SE}[\tilde \tau(B)] = \Lambda_0(B) \sqrt{u_0(B) \left ( \frac{1}{p \, \Lambda_0(B)} \right ) + u_1(A) \left (\frac{1 - p}{p \, N_1(A)} + \frac{q(B) \, q(A \backslash B) \, N_1(A)}{p \, \Lambda_0(B)^2} \right ) } . \]
Substituting \(\tilde \Lambda_0(B)\), \(\hat{p}\), \(\hat q(B)\), and
\(\hat q(A \backslash B)\) for \(\Lambda_0(B)\), \(p\), \(q(B)\), and
\(q(A \backslash B)\) yields a plug-in estimate \[
\tilde{\text{SE}}[\tilde \tau(B)] = \dfrac{N_1(A) \, n_0(B)}{n_1(A)} \sqrt{ \hat u_0(B) \left ( \dfrac{1}{n_0(B)} \right ) + u_1(A) \left (\frac{1}{n_1(A)} - \frac{1}{N_1(A)} + \dfrac{n_1(B) \, n_1(A \backslash B)}{n_1(A) \, n_0(B)^2} \right ) } . \]

\subsubsection{7.6 Hierarchical Model}\label{hierarchical-model}

We consider two alternatives to model 1. Unlike Sections 7.4 and 7.5, we
use a hierarchical model in which the thinning probabilities vary across
partitions according to a known family of parametric distributions. We
assume the probabilities are constant within partition.

Let \(\{B_i\}_{i=1}^{I}\) be an arbitrary partition of \(A\) into
disjoint partition elements such that \(A = \bigcup_{i=1}^{I} B_i\).
Model 4 is \begin{align*}
n_0(B_i) \sim & \ \text{Poisson} \left ( p(B_i) \, \Lambda_0(B_i) \right ) \\
\left [ n_1(B_1), \, \ldots, \, n_{1}(B_I), \, N_1(A) - n_1(A) \right ] \, | \, N_1(A) \sim & \ \text{Multinomial} \left (N_1(A), \, \left [ r(B_1), \, \ldots, \, r(B_I), \, 1- \sum_{i=1}^I r(B_i) \right ]  \right ) \\
\text{logit} \, p(B_i) \sim & \ \text{Normal}(\mu_p, \, \sigma_p^2) \\
\text{log} \, \Lambda_0(B_i)  \sim & \ \text{Normal}(\mu_0, \, \sigma_0^2) \\
\text{log} \, \Lambda_1(B_i)  \sim & \ \text{Normal}(\mu_1, \, \sigma_1^2)
\end{align*} where
\[r(B_i) = \frac{p(B_i) \, \Lambda_1(B_i)}{\Lambda_1(A)} .\]

Model 5 is \begin{align*}
n_0(B_i) \sim & \ \text{Negative Binomial} \bigl (p(B_i) \, \Lambda_0(B_i), \ u_{0}(B_i) \, p(B_i) \, \Lambda_0(B_i)  \bigr ) \\ 
\bigl [ n_1(B_1), \, \ldots, \, n_1(B_I), \, N_1(A) - n_1(A) \, \bigl ] \, | \, N_1(A) \sim & \ \text{MNH} \left ( N_1(A) - 1, \ \left [ s(B_1), \, \ldots, \ s(B_I), \ 1 - \sum_{i=1}^I s(B_i) \right ] \right ) \\
\text{logit} \, p(B_i) \sim & \ \text{Normal}(\mu_p, \, \sigma_p^2) \\
\text{log} \, \Lambda_0(B_i)  \sim & \ \text{Normal}(\mu_0, \, \sigma_0^2) \\
\text{log} \, \Lambda_1(B_i)  \sim & \ \text{Normal}(\mu_1, \, \sigma_1^2)
\end{align*} where
\[u_0(B_i) = \frac{\Lambda_0(B_i) - M(B_i)}{M(B_i)} \qquad \text{and} \qquad s(B_i) = p(B_i) \, M(B_i) = \frac{p(B_i) \, \Lambda_1(B_i) \, M(A)}{\Lambda_1(A)} .\]
As in Section 7.5, MNH denotes the multivariate negative hypergeometric
distribution. We define the negative binomial and negative
hypergeometric distributions as the total number of trials required to
get a predetermined number of successes when sampling with and without
replacement, respectively.

The parameter \(\mu_1\) is not identified because \(r\) and \(s\) are
invariant to scale. We therefore set \(\mu_1 = 0\) in the following
\texttt{Stan} code for Model 4 and 5 without loss of generality.

\begin{Shaded}
\begin{Highlighting}[]
\CommentTok{// Model 4 Code}
\KeywordTok{data}\NormalTok{ \{}
  \DataTypeTok{int}\NormalTok{\textless{}}\KeywordTok{lower}\NormalTok{=}\DecValTok{1}\NormalTok{\textgreater{} K;            }\CommentTok{// number of subregions}
  \DataTypeTok{array}\NormalTok{[K] }\DataTypeTok{int}\NormalTok{\textless{}}\KeywordTok{lower}\NormalTok{=}\DecValTok{0}\NormalTok{\textgreater{} n0;  }\CommentTok{// number of vendors without permits in each subregion}
  \DataTypeTok{array}\NormalTok{[K] }\DataTypeTok{int}\NormalTok{\textless{}}\KeywordTok{lower}\NormalTok{=}\DecValTok{0}\NormalTok{\textgreater{} n1;  }\CommentTok{// number of vendors with permits in each subregion}
  \DataTypeTok{int}\NormalTok{\textless{}}\KeywordTok{lower}\NormalTok{=}\DecValTok{0}\NormalTok{\textgreater{} N1;           }\CommentTok{// known number of permits/licenses citywide}
\NormalTok{\}}

\KeywordTok{parameters}\NormalTok{ \{}
  \DataTypeTok{real}\NormalTok{ mu\_p;}
  \DataTypeTok{real}\NormalTok{\textless{}}\KeywordTok{lower}\NormalTok{=}\DecValTok{0}\NormalTok{\textgreater{} sigma\_p;}
  \DataTypeTok{vector}\NormalTok{[K] alpha\_raw;}
  \DataTypeTok{real}\NormalTok{ mu\_0;}
  \DataTypeTok{real}\NormalTok{\textless{}}\KeywordTok{lower}\NormalTok{=}\DecValTok{0}\NormalTok{\textgreater{} sigma\_0;}
  \DataTypeTok{vector}\NormalTok{[K] eta0\_raw;}
  \DataTypeTok{real}\NormalTok{\textless{}}\KeywordTok{lower}\NormalTok{=}\DecValTok{0}\NormalTok{\textgreater{} sigma\_1;}
  \DataTypeTok{vector}\NormalTok{[K}\DecValTok{{-}1}\NormalTok{] eta1\_raw;}
\NormalTok{\}}

\KeywordTok{transformed parameters}\NormalTok{ \{}
  \DataTypeTok{vector}\NormalTok{[K] alpha = mu\_p + sigma\_p * alpha\_raw;}
  \DataTypeTok{vector}\NormalTok{\textless{}}\KeywordTok{lower}\NormalTok{=}\DecValTok{0}\NormalTok{, }\KeywordTok{upper}\NormalTok{=}\DecValTok{1}\NormalTok{\textgreater{}[K] p = inv\_logit(alpha);}
  \DataTypeTok{vector}\NormalTok{[K] eta0 = mu\_0 + sigma\_0 * eta0\_raw;}
  \DataTypeTok{vector}\NormalTok{\textless{}}\KeywordTok{lower}\NormalTok{=}\DecValTok{0}\NormalTok{\textgreater{}[K] lambda0 = exp(eta0);}
  \DataTypeTok{vector}\NormalTok{[K] eta1;}
\NormalTok{  eta1[}\DecValTok{1}\NormalTok{:(K}\DecValTok{{-}1}\NormalTok{)] = sigma\_1 * eta1\_raw;}
\NormalTok{  eta1[K] = {-}sum(eta1[}\DecValTok{1}\NormalTok{:(K}\DecValTok{{-}1}\NormalTok{)]);}
  \DataTypeTok{simplex}\NormalTok{[K] r = softmax(eta1);                      }
\NormalTok{\}}

\KeywordTok{model}\NormalTok{ \{}
\NormalTok{  alpha\_raw \textasciitilde{} normal(}\DecValTok{0}\NormalTok{,}\DecValTok{1}\NormalTok{);}
\NormalTok{  eta0\_raw \textasciitilde{} normal(}\DecValTok{0}\NormalTok{,}\DecValTok{1}\NormalTok{);}
\NormalTok{  eta1\_raw \textasciitilde{} normal(}\DecValTok{0}\NormalTok{, }\DecValTok{1}\NormalTok{);}
  
  \ControlFlowTok{for}\NormalTok{ (i }\ControlFlowTok{in} \DecValTok{1}\NormalTok{:K)}
\NormalTok{    n0[i] \textasciitilde{} poisson(p[i] * lambda0[i]);}
  
  \DataTypeTok{vector}\NormalTok{[K+}\DecValTok{1}\NormalTok{] theta;}
\NormalTok{  theta[}\DecValTok{1}\NormalTok{:K] = p .* r;}
\NormalTok{  theta[K+}\DecValTok{1}\NormalTok{] = }\DecValTok{1}\NormalTok{ {-} sum(p .* r);}

  \DataTypeTok{array}\NormalTok{[K+}\DecValTok{1}\NormalTok{] }\DataTypeTok{int}\NormalTok{ n1\_aug;}
\NormalTok{  n1\_aug[}\DecValTok{1}\NormalTok{:K] = n1;}
\NormalTok{  n1\_aug[K+}\DecValTok{1}\NormalTok{] = N1 {-} sum(n1);}

  \KeywordTok{target +=}\NormalTok{ multinomial\_lpmf(n1\_aug | theta);}
\NormalTok{\}}
\end{Highlighting}
\end{Shaded}

\begin{Shaded}
\begin{Highlighting}[]
\CommentTok{// Model 5 Code}
\KeywordTok{functions}\NormalTok{ \{}
  \DataTypeTok{real}\NormalTok{ nb\_lpmf(}\DataTypeTok{int}\NormalTok{ n, }\DataTypeTok{real}\NormalTok{ mu, }\DataTypeTok{real}\NormalTok{ rho) \{}
    \DataTypeTok{real}\NormalTok{ r = mu / rho;}
    \DataTypeTok{real}\NormalTok{ q = }\FloatTok{1.0}\NormalTok{ {-} }\FloatTok{1.0}\NormalTok{ / rho;}
    \ControlFlowTok{if}\NormalTok{ (r \textless{}= }\DecValTok{0}\NormalTok{ || q \textless{}= }\DecValTok{0}\NormalTok{ || q \textgreater{}= }\DecValTok{1}\NormalTok{) }\ControlFlowTok{return}\NormalTok{ negative\_infinity();}
    \ControlFlowTok{if}\NormalTok{ (n {-} r + }\DecValTok{1}\NormalTok{ \textless{}= }\DecValTok{0}\NormalTok{)             }\ControlFlowTok{return}\NormalTok{ negative\_infinity();}
    \ControlFlowTok{return}\NormalTok{ lgamma(n) {-} lgamma(r) {-} lgamma(n {-} r + }\DecValTok{1}\NormalTok{)}
\NormalTok{         + (n {-} r) * log(q) + r * log1m(q);}
\NormalTok{  \}}
  \DataTypeTok{real}\NormalTok{ mnh\_lpmf(}\DataTypeTok{array}\NormalTok{[] }\DataTypeTok{int}\NormalTok{ n\_aug, }\DataTypeTok{vector}\NormalTok{ r) \{}
    \DataTypeTok{int}\NormalTok{ J = size(n\_aug);}
    \DataTypeTok{int}\NormalTok{ N = }\DecValTok{0}\NormalTok{; }\ControlFlowTok{for}\NormalTok{ (j }\ControlFlowTok{in} \DecValTok{1}\NormalTok{:J) N += n\_aug[j];}
    \DataTypeTok{real}\NormalTok{ R = sum(r);}
    \ControlFlowTok{if}\NormalTok{ (R \textless{}= }\DecValTok{0}\NormalTok{ || N {-} R + }\DecValTok{1}\NormalTok{ \textless{}= }\DecValTok{0}\NormalTok{) }\ControlFlowTok{return}\NormalTok{ negative\_infinity();}
    \DataTypeTok{real}\NormalTok{ lp = }\DecValTok{0}\NormalTok{;}
    \ControlFlowTok{for}\NormalTok{ (j }\ControlFlowTok{in} \DecValTok{1}\NormalTok{:J) \{}
      \ControlFlowTok{if}\NormalTok{ (n\_aug[j] \textless{}= }\DecValTok{0}\NormalTok{ || r[j] \textless{}= }\DecValTok{0}\NormalTok{) }\ControlFlowTok{return}\NormalTok{ negative\_infinity();}
      \ControlFlowTok{if}\NormalTok{ (n\_aug[j] {-} r[j] + }\DecValTok{1}\NormalTok{ \textless{}= }\DecValTok{0}\NormalTok{)   }\ControlFlowTok{return}\NormalTok{ negative\_infinity();}
\NormalTok{      lp += lgamma(n\_aug[j]) {-} lgamma(r[j]) {-} lgamma(n\_aug[j] {-} r[j] + }\DecValTok{1}\NormalTok{);}
\NormalTok{    \}}
\NormalTok{    lp {-}= (lgamma(N) {-} lgamma(R) {-} lgamma(N {-} R + }\DecValTok{1}\NormalTok{));}
    \ControlFlowTok{return}\NormalTok{ lp;}
\NormalTok{  \}}
\NormalTok{\}}

\KeywordTok{data}\NormalTok{ \{}
  \DataTypeTok{int}\NormalTok{\textless{}}\KeywordTok{lower}\NormalTok{=}\DecValTok{1}\NormalTok{\textgreater{} K;            }\CommentTok{// number of subregions}
  \DataTypeTok{array}\NormalTok{[K] }\DataTypeTok{int}\NormalTok{\textless{}}\KeywordTok{lower}\NormalTok{=}\DecValTok{0}\NormalTok{\textgreater{} n0;  }\CommentTok{// number of vendors without permits in each subregion}
  \DataTypeTok{array}\NormalTok{[K] }\DataTypeTok{int}\NormalTok{\textless{}}\KeywordTok{lower}\NormalTok{=}\DecValTok{0}\NormalTok{\textgreater{} n1;  }\CommentTok{// number of vendors with permits in each subregion}
  \DataTypeTok{int}\NormalTok{\textless{}}\KeywordTok{lower}\NormalTok{=}\DecValTok{0}\NormalTok{\textgreater{} N1;           }\CommentTok{// known number of permits/licenses citywide}
  \DataTypeTok{vector}\NormalTok{\textless{}}\KeywordTok{lower}\NormalTok{=}\DecValTok{1}\NormalTok{\textgreater{}[K] rho;    }\CommentTok{// expected proportion of respondents per market}
\NormalTok{\}}

\KeywordTok{parameters}\NormalTok{ \{}
  \DataTypeTok{real}\NormalTok{ mu\_p;}
  \DataTypeTok{real}\NormalTok{\textless{}}\KeywordTok{lower}\NormalTok{=}\DecValTok{0}\NormalTok{\textgreater{} sigma\_p;}
  \DataTypeTok{vector}\NormalTok{[K] alpha\_raw;}
  \DataTypeTok{real}\NormalTok{ mu\_0;}
  \DataTypeTok{real}\NormalTok{\textless{}}\KeywordTok{lower}\NormalTok{=}\DecValTok{0}\NormalTok{\textgreater{} sigma\_0;}
  \DataTypeTok{vector}\NormalTok{[K] eta0\_raw;}
  \DataTypeTok{real}\NormalTok{\textless{}}\KeywordTok{lower}\NormalTok{=}\DecValTok{0}\NormalTok{\textgreater{} sigma\_1;}
  \DataTypeTok{vector}\NormalTok{[K] eta1\_raw;}
\NormalTok{\}}

\KeywordTok{transformed parameters}\NormalTok{ \{}
  \DataTypeTok{vector}\NormalTok{[K] alpha = mu\_p + sigma\_p * alpha\_raw;}
  \DataTypeTok{vector}\NormalTok{\textless{}}\KeywordTok{lower}\NormalTok{=}\DecValTok{0}\NormalTok{, }\KeywordTok{upper}\NormalTok{=}\DecValTok{1}\NormalTok{\textgreater{}[K] p = inv\_logit(alpha);}
  \DataTypeTok{vector}\NormalTok{[K] eta0 = mu\_0 + sigma\_0 * eta0\_raw;}
  \DataTypeTok{vector}\NormalTok{\textless{}}\KeywordTok{lower}\NormalTok{=}\DecValTok{0}\NormalTok{\textgreater{}[K] lambda0 = exp(eta0);}
  \DataTypeTok{vector}\NormalTok{\textless{}}\KeywordTok{lower}\NormalTok{=}\DecValTok{0}\NormalTok{\textgreater{}[K] mu0 = p .* lambda0;}
  \DataTypeTok{vector}\NormalTok{\textless{}}\KeywordTok{lower}\NormalTok{=}\DecValTok{0}\NormalTok{\textgreater{}[K] M = mu0 ./ rho;}
  \DataTypeTok{vector}\NormalTok{[K] eta1 = sigma\_1 * eta1\_raw;}
  \DataTypeTok{vector}\NormalTok{\textless{}}\KeywordTok{lower}\NormalTok{=}\DecValTok{0}\NormalTok{\textgreater{}[K] lambda1 = exp(eta1);}
  \DataTypeTok{vector}\NormalTok{\textless{}}\KeywordTok{lower}\NormalTok{=}\DecValTok{0}\NormalTok{\textgreater{}[K+}\DecValTok{1}\NormalTok{] s;}
\NormalTok{  s[}\DecValTok{1}\NormalTok{:K] = sum(lambda0 ./ rho) * p .* (lambda1 / sum(lambda1));}
\NormalTok{  s[K+}\DecValTok{1}\NormalTok{] = sum(lambda0 ./ rho) * (}\DecValTok{1}\NormalTok{ {-} sum(p .* (lambda1 / sum(lambda1))));}
\NormalTok{\}}

\KeywordTok{model}\NormalTok{ \{}
\NormalTok{  alpha\_raw \textasciitilde{} normal(}\DecValTok{0}\NormalTok{, }\DecValTok{1}\NormalTok{);}
\NormalTok{  eta0\_raw  \textasciitilde{} normal(}\DecValTok{0}\NormalTok{, }\DecValTok{1}\NormalTok{);}
\NormalTok{  eta1\_raw  \textasciitilde{} normal(}\DecValTok{0}\NormalTok{, }\DecValTok{1}\NormalTok{);}

  \ControlFlowTok{for}\NormalTok{ (i }\ControlFlowTok{in} \DecValTok{1}\NormalTok{:K)}
    \KeywordTok{target +=}\NormalTok{ nb\_lpmf(n0[i] | mu0[i], rho[i]);}

  \DataTypeTok{array}\NormalTok{[K+}\DecValTok{1}\NormalTok{] }\DataTypeTok{int}\NormalTok{ n1\_aug;}
\NormalTok{  n1\_aug[}\DecValTok{1}\NormalTok{:K] = n1;}
\NormalTok{  n1\_aug[K+}\DecValTok{1}\NormalTok{] = N1 {-} sum(n1);}
  \KeywordTok{target +=}\NormalTok{ mnh\_lpmf(n1\_aug | s);}
\NormalTok{\}}
\end{Highlighting}
\end{Shaded}

\end{document}